%% file: paper.tex
\documentclass[acmtog,timestamp]{acmart}

\setcitestyle{square}

\input{preamble}

\setcopyright{rightsretained}
\acmDOI{10.475/123_4}

\acmISBN{123-4567-24-567/08/06}

\acmConference[]{ACM SIGGRAPH Conference}{Aug 2018}{Vancouver, Canada} 
\acmYear{2018}
\copyrightyear{2018}

\acmSubmissionID{***}

\begin{document}
\title{Palette-based image decomposition, harmonization, and color transfer}
% \titlenote{Produces the permission block, and
  % copyright information}
% \subtitlenote{The full version of the author's guide is available as
  % \texttt{acmart.pdf} document}

% \author{Anonymous}
\author{Jianchao Tan}
\affiliation{\institution{George Mason University}
}
\email{tanjianchaoustc@gmail.com}

\author{Jose Echevarria}
\affiliation{\institution{Adobe Research}
}
\email{echevarr@adobe.com}

\author{Yotam Gingold}
\affiliation{\institution{George Mason University}
}
\email{ygingold@gmu.edu }

\input{abstract}





\input{teaser}

\maketitle

\input{introduction}
\input{related}

\input{palette}
\input{harmonization}

\input{perceptual}
\input{video}
\input{transfer}
\input{conclusion}

\bibliographystyle{ACM-Reference-Format}
\bibliography{bib/timemap.bib,bib/singlelayer.bib,bib/pigmento.bib,bib/harmonization.bib}

\end{document}

%% file: preamble.tex
\usepackage{booktabs} % For formal tables

\renewcommand{\paragraph}[1]{{\vspace{2.05ex}\par\noindent\normalsize\sffamily\bfseries\kern-2.5pt#1\quad}}

\usepackage{graphicx}
\graphicspath{{./figs/}}

\usepackage[ruled,linesnumbered]{algorithm2e}

\usepackage{amsmath}
\usepackage{mathtools}
\usepackage{amssymb}
\usepackage{bm}
\usepackage{wrapfig}
\usepackage{hhline}
\usepackage{soul}
\usepackage{comment}

\usepackage[mathletters]{ucs}
\usepackage[utf8x,utf8]{inputenc}
\usepackage{upgreek}
\usepackage{dblfloatfix}

\usepackage{microtype}

\usepackage{booktabs}

\DeclareMathOperator*{\argmin}{arg\,min}

\newif\ifsubmit
\submittrue
\ifsubmit
    \newcommand{\todo}[1]{}
    \newcommand{\TODO}[1]{}
    
    \newcommand{\yotam}[1]{}
    \newcommand{\jose}[1]{}
    \newcommand{\jianchao}[1]{}
    
    \newcommand{\topic}[1]{}
    
    \newcommand{\rev}[1]{#1}
    \newcommand{\revbegin}{}
    \newcommand{\revend}{}

\else
    \newcommand{\todo}[1]{\textcolor{red}{TODO: #1}}
    \newcommand{\TODO}[1]{\textcolor{red}{\textbf{****** #1 ******}}}
    
    \newcommand{\yotam}[1]{{\color{purple}\textsc{Yotam:} #1}}
    \newcommand{\jose}[1]{{\color{orange}\textsc{Jose:} #1}}
    \newcommand{\jianchao}[1]{{\color{cyan}\textsc{Jianchao:} #1}}
    
    \newcommand{\topic}[1]{{\color{magenta}\textsc{Topic:} #1\\}}
    \newcommand{\rev}[1]{{\color{blue}#1}}
    \newcommand{\revbegin}{\color{blue}}
    \newcommand{\revend}{\color{black}}
\fi

\usepackage{letltxmacro}
\LetLtxMacro{\oldmarginpar}{\marginpar}
\renewcommand{\marginpar}[1]{\oldmarginpar{\color{red}\tiny #1}}

%% file: abstract.tex
\begin{abstract}
We present a palette-based framework for color composition for visual applications.
Color composition is a critical aspect of visual applications in art, design, and visualization.
The color wheel is often used to explain pleasing color combinations in geometric terms,
and, in digital design, to provide a user interface to visualize and manipulate colors.

We abstract relationships between palette colors as a compact set of axes describing \emph{harmonic templates} over perceptually uniform color wheels.
Our framework provides a basis for a variety of color-aware image operations, such as
color harmonization and color transfer, \rev{and can be applied to videos}.

To enable our approach, we introduce an extremely scalable and efficient yet simple palette-based image decomposition algorithm. Our approach is based on the geometry
of images in RGBXY-space. This new geometric approach is orders of magnitude more efficient than previous work and requires no numerical optimization.
\rev{We demonstrate a real-time layer decomposition tool. After preprocessing, our algorithm can decompose 6 MP images into layers in 20 milliseconds.}

\rev{We also conducted three large-scale, wide-ranging perceptual studies on the perception of harmonic colors and harmonization algorithms.}
\end{abstract}

%\jose{ABSTRACT1: Color carries a lot of semantic information. Thus, it is critical that any color-aware tool is able to model and preserve the high-level relationships between the main colors in an image, palette, theme or design. In this paper we propose a simple and intuitive parameterization of these relationships, which helps posing and solving previous problems and developing new tools easily and efficiently. Our parameterization leverages the classical harmonic templates from color composition theory to model relationships with respect to their positions around a color wheel, and their distances to the axes describing each template. By analyzing the structure and purpose of the different templates, we propose and demonstrate specific metrics for problems like palette transfer and enhancement, harmonization of images, or color-aware searches.}

%% file: teaser.tex
\begin{teaserfigure}
\centering
\includegraphics[width=\textwidth]{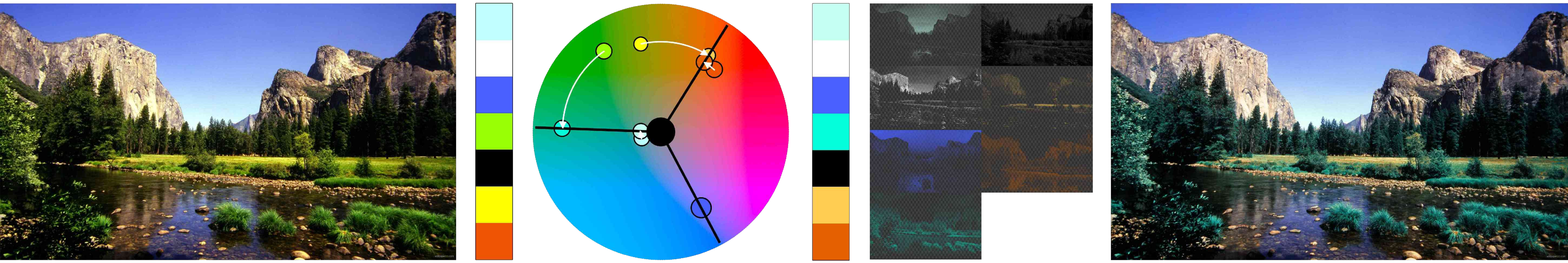}
\caption{Our palette-based color harmony template is able to express harmonization and various other color operations in a concise manner. Our edits make use of a new, extremely efficient image decomposition technique based on the 5D geometry of RGBXY-space.
\yotam{This example is too subtle. I think we should make some kind of overview figure.}
}
\label{fig:teaser}
\end{teaserfigure}

%% file: introduction.tex
\section{Introduction}
\label{sec:harmony_introduction}

Color composition is critical in visual applications in art, design, and visualization. Over the centuries, different theories about how colors interact with each other have been proposed~\cite{westland2007colour}. While it is arguable whether a universal and comprehensive color theory will ever exist, most previous proposals share in common the use of a color wheel (with hue parameterized by angle) to explain pleasing color combinations in geometric terms. In the digital world, the color wheel often serves as a user interface to visualize and manipulate colors. This has been explored in the literature for specific applications in design~\cite{AdobeColor2017} and image editing~\cite{Cohen-Or:2006:CH:1179352.1141933}.

In this paper, we embrace color wheels to present a new framework where color composition concepts are easy and intuitive to formulate, solve for, visualize, and interact with; for applications in art, design, or visualization.
Our approach is based on palettes and relies on palette-based image decompositions.
To fully realize it as a powerful image editing tool, we introduce an extremely efficient yet simple new image decomposition algorithm.
% \yotam{I think we should summarize the following before getting into specifics. Our approach is based on palettes not hue histograms, everything becomes easy to formulate, and we introduce a new, fast and simple image decomposition.}\jose{ok!}

%\marginpar{\yotam{Our color space (LCh wheel)}}
We define our color relationships in the CIE LCh color space (the cylindrical projection of CIE Lab). Contrary to previous work using HSV color wheels, the LCh color space ensures that perceptual effects are accounted for with no additional processing. For example, a simple global rotation of hue in LCh-space (but not HSV-space) preserves the perceived lightness or gradients in color themes and images.

%\marginpar{\yotam{Our model in this color space (palettes not histograms)}}
To represent color information, we adopt the powerful palette-oriented point of view~\cite{mellado2017constrained}
and propose to work with color palettes of arbitrary numbers of swatches. Unlike hue histograms, color palettes or swatches can come from a larger variety of sources (extracted from images, directly from user input, or from generative algorithms) and capture the 3D nature of LCh in a compact way. They provide intuitive interfaces and visualizations as well.

\begin{figure}
	\centering
	\includegraphics[width=\columnwidth]{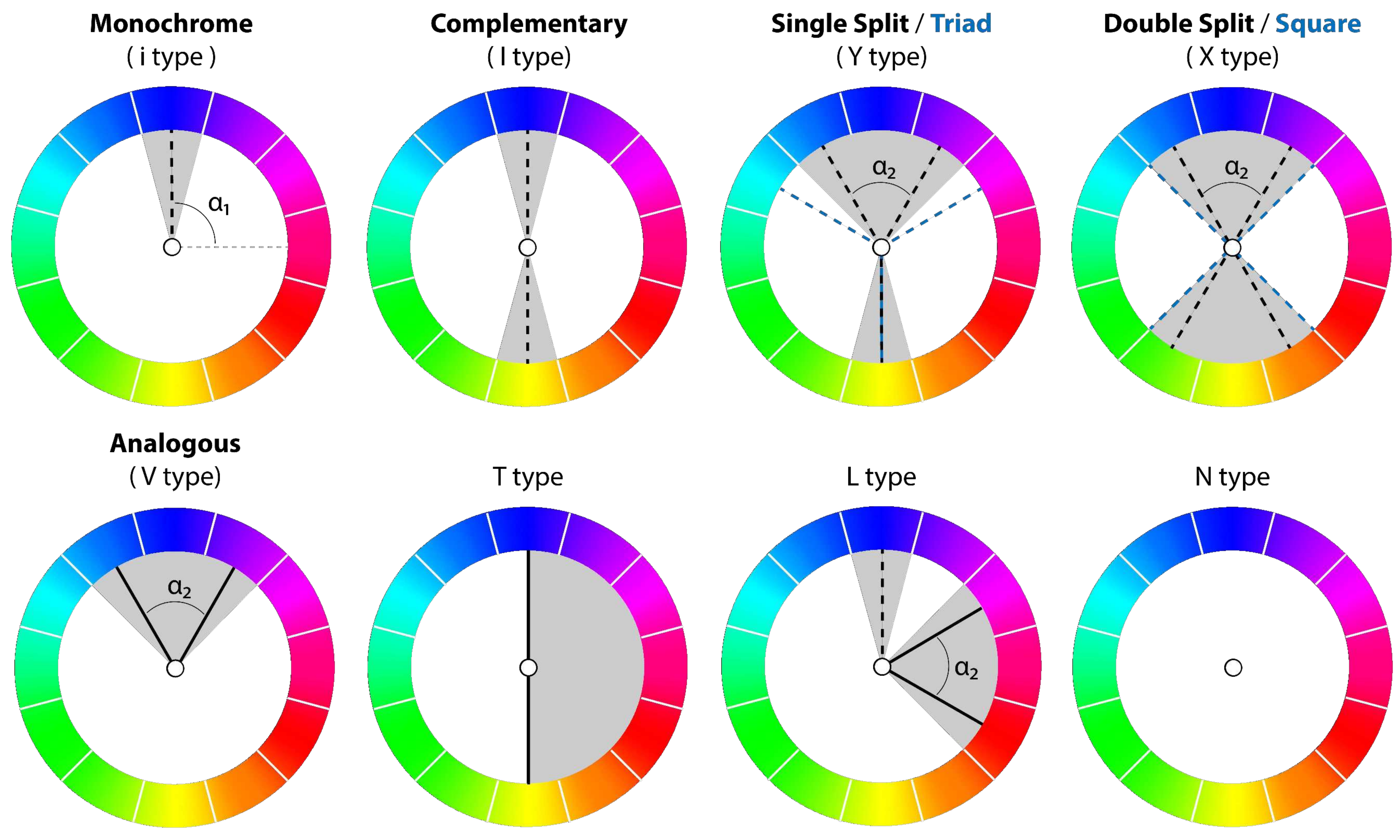}
	\caption{Comparison between the sector-based hue harmonic templates from~\cite{tokumaru2002color} (shaded in grey), and our new axis-based ones. We use two different types of axes: the dashed ones attract colors towards them; the solid ones define sectors between them, so colors inside remain in the same position, but colors outside are attracted towards them. We found this distinction helps handling templates like \emph{analogous} properly. Note that our templates derived from~\cite{itten1970elements} separate \emph{Y type} into \emph{single split} and \emph{triad}, and the same for \emph{X type}. These templates are popular among creatives, but they are also in agreement with the definitions of similarity and ambiguity by Moon and Spencer~\shortcite{Moon44}. Although we don't use it in our results, our approach can also describe hybrid templates like \emph{L type}. Each template can be modeled by a single global rotation $\alpha_1$, although some of them have a secondary degree of freedom $\alpha_2$ that enforces symmetry. In this paper we focus on \emph{monochrome}, \emph{complementary}, \emph{single split}, \emph{triad}, \emph{double split}, \emph{square} and \emph{analogous}.}
	\label{fig:template_axis}
\end{figure}

%\marginpar{\yotam{Our color harmonization/operators}}
Color palettes also simplify the modelling and formulation of relationships between colors. This last point enables the simplification of harmonic templates and other relationships into a set of a few 3D axes that capture color structure in a meaningful and compact way.
This is useful for various color-aware tasks. We demonstrate applications to color harmonization and color transfer.
Instead of using the sector-based templates from Matsuda~\cite{tokumaru2002color}
(appropriate for hue histograms) we derive our harmonic templates from classical color theory~\cite{itten1970elements,Birren1969} (see Figures~\ref{fig:template_axis} and~\ref{fig:LC_templates}). We also propose new color operations using this axes-based representation.
%\yotam{What are our new color operators? Warm/cool contrast; Taking into account less dominant colors.}
%
Our proposed framework can be used by other palette-based systems and workflows,
either for palette improvement or image editing.

At the core of our and other recent approaches~\cite{Chang:2015:PPR,Tan:2016:DIL,aksoy2017unmixing,zhang2017palette}
to image editing,
images are decomposed into a palette and associated per-pixel compositing or mixing parameters.
We propose a new, extremely efficient yet simple and robust algorithm to do so.
Our approach is inspired by the geometric palette extraction technique of Tan et al.~\shortcite{Tan:2016:DIL}. We consider the geometry of 5D RGBXY-space,
which captures color as well as spatial relationships
and eliminates numerical optimization.
After an initial palette is extracted (given an RMSE reconstruction threshold),
the user can edit the palette and obtain new decompositions instantaneously.
Our algorithm's performance is extremely efficient even for
very high resolution images ($\geq 100$ megapixels)---20x faster than the state-of-the-art \cite{aksoy2017unmixing}.
Working code is provided in Section~\ref{sec:harmony_Palette_Decompostion}.
Our algorithm is a key contribution which enables our approach and many other
applications proposed in the literature.
% explore different color transfer approaches, effortlessly formulated with our proposed framework. 
%\jose{Do we want to mention both methods here? If so, we may need to motivate the need for two methods}

%Image color transferring is one of popular photo editing tasks targeting all levels of expertise. Existing color transferring works can be classified into three categories: first are based on some global color histogram matching between two images; second are based on image palette extraction and image decomposition, palette color transferring will enable final image color transferring; third are based on deep learning techniques, these works usually learn a mapping function from a database and apply this function to an input image. Our color transferring is based on color harmonization template fitting, which is actually converted color transferring into a matching problem between two templates.

In summary, this papers makes the following contributions:
\begin{itemize}
	\item A new palette-based color harmonization framework, general enough to model classical harmonic relationships, new color composition operations, and a compact structure for other color-aware applications, \rev{also applicable to video}.
% 	\yotam{What precisely is new?}\jose{Our combination of palettes + axes}
	\item An extremely efficient, geometric approach for decomposing an image into spatially coherent additive mixing layers by analyzing the geometry of an image in RGBXY-space. Its performance is virtually independent from the size of the image or palette. Our decomposition can be \emph{re-computed} instantaneously for a new RGB palette, allowing designers to edit the decomposition in real-time.
	% The other builds on previous optimization approaches, with a new more efficient quadratic energy function.
	\item \rev{Three large-scale, wide-ranging perceptual studies on the perception of harmonic colors and our algorithm.}
\end{itemize}
We demonstrate other applications like color transfer, greatly simplified by our framework.

%Our work has three contributions. 
%First, we provide two new new methods for extracting additive mixing layers from input image, one is based on optimization of a quadratic energy function, the other is based on RGBXY convex hull geometry, latter's performance is not depend on image size, which is efficient for large size image. 
%Second, we explored the palette based image harmonization and image contrast adjustment, which are user friendly and more intuitive than previous all pixels colors' harmonization. 
%Finally, we proposed a novel image color transfer method by using palette harmonization template fitting result as intermediate tool.

%% We don't need this paragraph, except for ourselves.
% The rest of the paper is organized as follows. Section~\ref{sec:harmony_relatedwork} gives a review of related works in terms of different part of our pipeline. In Section~\ref{sec:harmony_Palette_Decompostion}, we introduce how to extract palette from image and two methods for decomposing image into a set of additive mixing layers.  In Section~\ref{sec:harmony_Harmonization}, we introduce a palette color harmonization template fitting method to enable image color harmonization \jose{Shall we put harmonization first?}. Section~\ref{sec:harmony_ColorTransfer} introduces a novel color transferring method that is based on palette harmonization template fitting. Final section gives conclusion, limitation discussion and future work.

%% file: related.tex
\section{Related Work}
\label{sec:harmony_relatedwork}
There are many works related with our contributions and their applications. In the following we cover the most relevant ones.

\subsection{Color Harmonization}
\label{sec:harmony_relatedwork:harmonization}
Many existing works have applied different concepts from traditional color theory for artists to improve the color composition of digital images. In their seminal paper, \citet{Cohen-Or:2006:CH:1179352.1141933} use hue histograms and harmonic templates defined as sectors of hue-saturation in HSV color space~\cite{tokumaru2002color}, to model and manipulate color relationships. They fit a template (optimal or arbitrary) over the image histogram, so they can shift hues accordingly to harmonize colors or composites from several sources. Additional processing is needed to ensure spatial smoothness. Several people have built on top of this work, extending or improving parts of their proposed framework. \citet{sm_vidHarm_icvgip_08} extended it to video, focusing on temporal coherence between successive frames. Improvements to the original fitting have been proposed based on the number of pixels for each HSV value~\cite{huo2009improved}, the visual saliency~\cite{baveye2013saliency}, the extension and visual weight of each color.~\cite{baveye2013saliency}, or geodesic distances~\cite{li2015image}. Tang et al.~\shortcite{tang2010image} improves some artifacts during the recoloring of \cite{Cohen-Or:2006:CH:1179352.1141933}. Chamaret et al.~\shortcite{chamaret2014harmony} defines and visualizes a per-pixel harmony measure to guide interactive user edits.  

Instead of using hue histograms from images, our framework is built on top of color palettes, independently of their source. Given the higher level of abstraction of palettes, we simplify harmonic templates to arrangements of axes in chroma-hue space (from LCh), interpreted and derived directly from classical color theory~\cite{itten1970elements,Birren1969}. This more general and simpler representation makes for more intuitive metrics, easier to solve, that enable a wider range of applications. When working with images, this approach fits perfectly with our proposed palette extraction and image decomposition for very efficient and robust image recoloring. Related to our approach, \citet{mellado2017constrained} is also able to pose harmonization as a set of constrains within their general constrained optimization framework. Our new templates, posed in LCh space, could be added as additional constraints.
Finally, there is a different definition for harmony in composited images, in terms of contrast, texture, noise or blur. Works dealing with it~\cite{sunkavalli2010multi,2017arXiv170300069T} focus on a completely different set of goals and challenges than the work discussed above.

%There are two other type image harmonization works. One is sunkavalli et al.~\cite{sunkavalli2010multi}, which use a multi-scale technique to transfer the appearance of one image to another, matching contrast, texture, noise, and blur and avoiding image artifacts by manipulating the scales of a pyramid decomposition of image. The other is Tsai et al.~\cite{2017arXiv170300069T}, which use deep learning technique to learn the mapping from input image to a harmonized image. They are different from standard template fitting procedure as discussed above. 

\subsection{Palette Extraction and Image Decomposition}
\label{sec:harmony_relatedwork:palette_decomposition}
\paragraph{Palette Extraction}
%Because it level of abstraction, palette extraction is nowadays the first step of many pipelines for image processing including recoloring, classification or enhancement. 
A straightforward approach consists of using a k-means method to cluster the existing colors in an image, in RGB space~\cite{Chang:2015:PPR,phan2017color,zhang2017palette}. A different approach consists of computing and simplifying the convex hull enclosing all the color samples~\cite{Tan:2016:DIL}, which provides more general palettes that better represent the existing color gamut of the image.
A similar observation was made in the domain of hyperspectral image unmixing \cite{craig1994minimum}. (With hyperspectral images, palette sizes are smaller than the number of channels, so the problem is one of fitting a minimum-volume simplex around the colors. The vertices of a high-dimensional simplex become a convex hull when the data is projected to lower dimensions.)
Morse et al.~\shortcite{morse2007image} work in HSL space, using a histogram to find the dominant hues, then to find shades and tints within them. Human perception has also been taken into account in other works, training regression models on crowd-sourced datasets.~\cite{o2011color,lin2013modeling}. Some physically-based approaches try to extract wavelength-dependent parameters to model the original pigments used paintings.~\cite{tan2017pigmento,aharoni2017pigment}. Our work builds on top of Tan et al.~\shortcite{Tan:2016:DIL}, adding a fixed reconstruction error threshold for automatic extraction of palettes of optimal size, as described in Section~\ref{sec:harmony_Palette_Decompostion:palette}.

\paragraph{Image Decomposition}
For recoloring applications, it is also critical to find a mapping between the extracted color palette and the image pixels. Recent work is able to decompose the input image into separate layers according to a palette. Tan et al.~\shortcite{Tan:2016:DIL} extract a set of ordered translucent RGBA layers, based on a optimization over the standard alpha blending model. Order-independent decompositions can be achieved using additive color mixing models~\cite{aksoy2017unmixing,lin2017layer,zhang2017palette}. For the physically-based palette extraction methods mentioned previously~\cite{tan2017pigmento,aharoni2017pigment}, layers correspond to the extracted multi-spectral pigments.
% \citet{Chang:2015:PPR} support palette-based
% color manipulation of images by transferring a palette color's change to pixels based on its estimated influence; no complete image decomposition is performed.
We prefer a full decomposition to a (palette-based) edit transfer approach like
\citet{Chang:2015:PPR}'s. With a full decomposition, edits are trivial to apply and spatial edits become possible (though we do not explore spatial edits in this work).
We present a new, efficient method for layer decomposition, based on the additive color mixing model  (Section~\ref{sec:harmony_Palette_Decompostion:RGBXYconvexhull}).
Our approach leverages 5D RGBXY-space geometry to enforce spatial smoothness on the layers. This geometric approach is significantly more efficient than previous approaches
in the literature, easily handling images up to 100 megapixels in size.

\subsection{Color Transfer}
\label{sec:harmony_relatedwork:transfer}
We also explore color transfer as an \emph{application} of our work.
%While it is not the main purpose of our framework, we also explore color transfer with it.
Color transfer is a vast field with contributions from the vision and graphics communities. As such, we describe only the most closely related work to our approach. 
% Palette based transfer
Hou et al.~\shortcite{hou2007color} conceptualize and apply color themes as hue histograms in HSV space. 
Wang et al.~\shortcite{wang2010data} solve an optimization that simultaneously considers a desired color theme, texture-color relationships as well as automatic or user-specified color constraints. 
Phan et al.~\shortcite{phan2017color} explored the order of colors within palettes to establish correspondences and enable interpolation. 
Nguyen et al.~\shortcite{nguyen2017group} find a group color theme from multiple palettes from multiple images using a modified k-means clustering method, and use it to recolor all the images in a consistent way.
Han et al.~\shortcite{Han2013,han2017cartoon} compute a distance metric between palettes in the \emph{color mood} space, and then sort and match colors from palettes according to their brightness.  
Munshi et al.~\shortcite{munshi2015palette} match colors between palettes according to their distance in Lab space. Based on our harmonic templates, palettes, and the LCh color space; we propose several intuitive metrics for color transfer that take into account human perception for goals like colorfulness, preservation of original colors, or harmonic composition. The final image recoloring is performed using our layer decomposition.

%% file: palette.tex
\section{Palette extraction and image decomposition}
\label{sec:harmony_Palette_Decompostion}

A good palette for image editing is one that closely captures the underlying colors the image was made with (or could have been made with), even if those colors do not appear in their purest form in the image itself.
\citet{Tan:2016:DIL} observed that the color distributions from paintings and natural images take on a convex shape in RGB space. As a result, they proposed to compute the convex hull of the pixel colors. The convex hull tightly wraps the observed colors. Its vertex colors can be blended with convex weights (positive and summing to one) to obtain any color in the image. The convex hull may be overly complex, so they propose an iterative simplification scheme to a user-desired palette size. After simplification, the vertices become a palette that represents the colors in the image.
%\jose{This motivation feels a bit long, specially since we build on top of Tan et al 2016}
%\yotam{I shortened it. I moved the hyperspectral imaging stuff away.}
%We believe this approach produces palettes that capture better the relationships between the colors in the image, so

% We extend \citet{Tan:2016:DIL} in two ways: automatic palette size selection and decomposition into additive mixing layers.
% extraction and novel image layer decomposition methods.

We extend \citet{Tan:2016:DIL}'s work in two ways. First, we propose a simple, geometric layer decomposition method that is orders of magnitude more efficient than the state-of-the-art.
Working code for our entire decomposition algorithm can be written in under \rev{50} lines (Figure~\ref{fig:decomposition_algorithm}).
Second, we propose a simple scheme for automatic palette size selection.

% Automatic Palette Size Selection
\subsection{Palette Extraction}
\label{sec:harmony_Palette_Decompostion:palette}
In \citet{Tan:2016:DIL}, the convex hull of all pixel colors is computed and then simplified to a user-chosen palette size.
To summarize their approach, the convex hull is simplified greedily as a sequence of constrained edge collapses \cite{Garland:1997:SSU}.
An edge is collapsed to a point constrained to strictly add volume \cite{Sander:2000:SC} while minimizing the distance to its incident faces.
The edge whose collapse adds the least overall volume is chosen next, greedily.
After each edge is collapsed, the convex hull is recomputed, since the new vertex could indirectly cause other vertices to become concave (and therefore redundant).
Finally, simplification may result in out-of-gamut colors, or points that lie outside the RGB cube. As a final step, \citet{Tan:2016:DIL} project all such points to the closest point on the RGB cube. This is the source of reconstruction error in their approach;
some pixels now lie outside the simplified convex hull and cannot be reconstructed.

We improve upon this procedure with the observation that the reconstruction error can be measured geometrically, even before layer decomposition, as the RMSE of every pixel's distance to the simplified convex hull. (Inside pixels naturally have distance 0.) Therefore, we propose a simple automatic palette size selection based on a user-provided RMSE reconstruction error tolerance ($\frac{2}{255}$ in our experiments).
For efficiency, we divide RGB-space into $32 \times 32 \times 32$ bins (a total of $2^{15}$ bins). We measure the distance from each non-empty bin to the simplified convex hull, weighted by the bin count. We start measuring the reconstruction error once the number of vertices has been simplified to 10.
By doing this, we are able to obtain palettes with an optimal number of colors automatically. This removes the need for the user to choose the palette size manually,
leading to better layer decompositions.

\yotam{The following is speculative and could be removed. I think it's a neat idea, though.}
(If non-constant palette colors were acceptable, instead of clipping one could cast a ray from each pixel towards the out-of-gamut vertex; the intersection of the ray with the RGB cube would be the palette color for that pixel. There would be zero reconstruction error. The stopping criteria could be the non-uniformity of a palette color, measured by the area of the RGB cube surface intersected with the simplified convex hull itself.)

\begin{figure}
	\centering
	\includegraphics[width=\columnwidth]{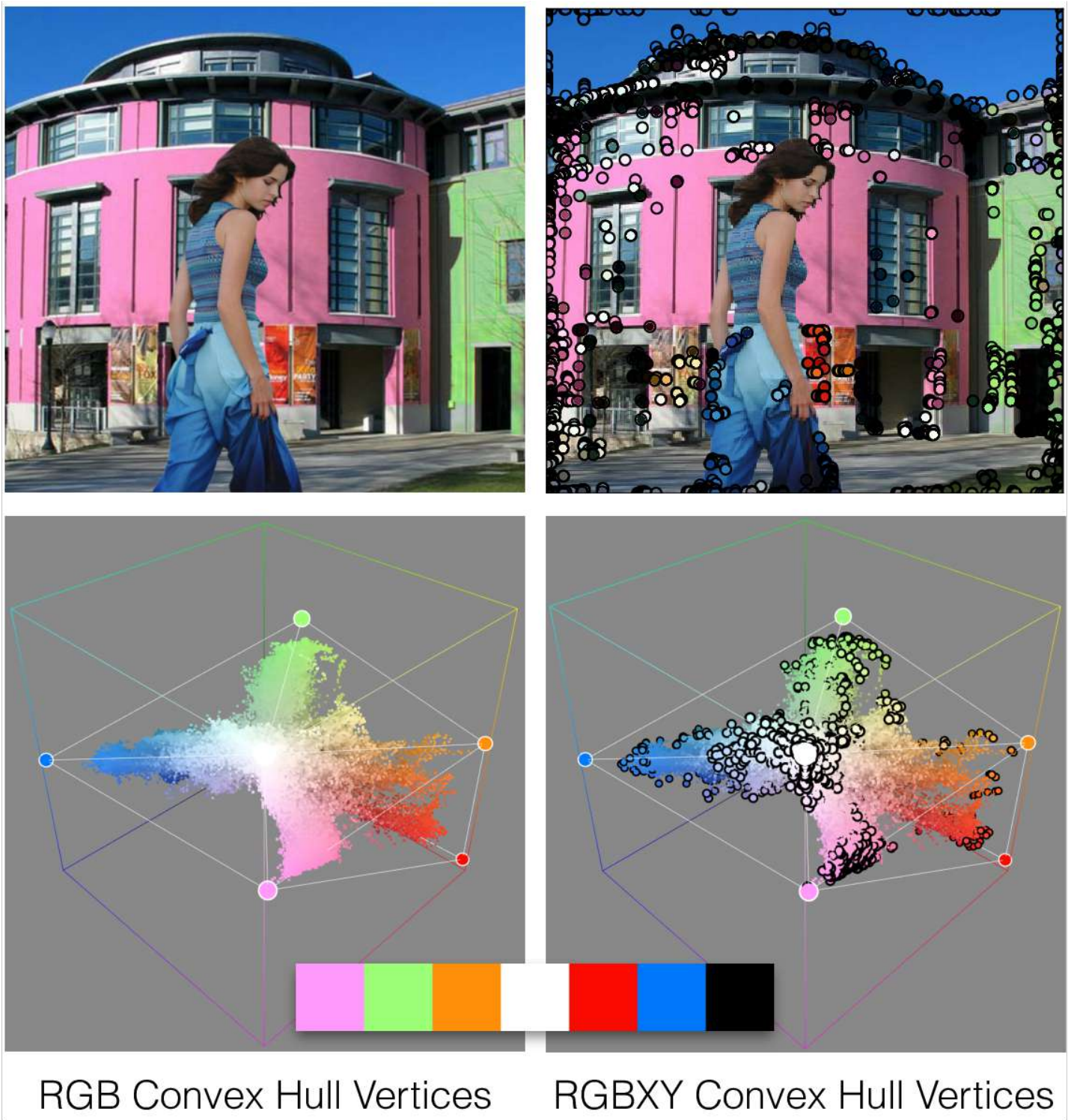}
	\caption{Visualization of the two convex hulls. \emph{Left}: the simplified RGB convex hull is the basis for the methods in Tan et al.~\shortcite{Tan:2016:DIL}, capturing the colors of an image but not their spatial relationships. \emph{Right}: Our 5D RGBXY convex hull captures color and spatial relationship at the same time. We visualize its vertices as small circles; its 5D simplices are difficult to visualize. Our approach splits image decomposition into a two-level geometric problem. The first level are the RGBXY convex hull vertices that mix to produce any pixel in the image. The second level are the simplified RGB convex hull vertices, which serve as the palette RGB colors. Since the RGBXY convex hull vertices lie inside the RGB convex hull, we find mixing weights that control the color of the RGBXY vertices. The two levels combined allow instant recoloring of the whole image. The top right image shows the locations of the RGBXY vertices in image space. The bottom row shows the geometric relationships between the 3D and 5D convex hull vertices, and how the simplified RGB convex hull captures the same color palette for both algorithms.}
	%\caption{Our two proposed methods. Both methods are using same fixed palette from previous convex hull simplification step. For quadratic optimization, we directly solve the mixing weights map from whole image pixels colors (bottom left). For RGBXY convex hull method, we decompose the problem into two level convex hull geometry problem, first level is RGBXY convex hull vertices mixing weights for whole image pixels, and second level is palette mixing weights for these RGBXY convex hull vertices. Then merge two level mixing weights together to be palette mixing weights for whole image pixels. The right top image shows the locations of these RGBXY convex hull vertices in image.}
	\label{fig:RGBXY_method_illustration}
\end{figure}

\subsection{Image decomposition via RGBXY convex hull}
\label{sec:harmony_Palette_Decompostion:RGBXYconvexhull}

From their extracted palettes, \citet{Tan:2016:DIL} solved a non-linear optimization problem to decompose an image into a set of ordered, translucent RGBA layers suitable for the standard ``over'' compositing operation.
While this decomposition is widely applicable (owing to the ubiquity of ``over'' compositing), the optimization is quite lengthy due to the recursive nature of the compositing operation, which manifests as a polynomial whose degree is the palette size.
Others have instead opted for additive mixing layers \cite{aksoy2017unmixing,lin2017layer,zhang2017palette} due to their simplicity.
A pixel's color is a weighted sum of the palette colors.

In this work, we adopt linear mixing layers as well. We provide a fast and simple, yet spatially coherent, geometric construction.

Any point $\mathbf{p}$ inside a simplex (a triangle in 3D, a tetrahedron in 3D, etc.) has a unique set of barycentric coordinates, or convex additive mixing weights such that
$
    \mathbf{p} = \sum_i w_i \mathbf{c}_i,
$
where the mixing weights $w_i$ are positive and sum to one,
and $\mathbf{c}_i$ are the vertices of the simplex.
In our setting, the simplified convex hull is typically not a simplex, because the palette has more than 4 colors. There still exist convex weights $w_i$ for arbitrary polyhedron,
known as generalized barycentric coordinates \cite{floater2015generalized}, but they are typically non-unique.
A straightforward technique to find generalized barycentric coordinates is to first compute a tessellation of the polyhedron (in our case, the simplified convex hull)
into a collection of non-overlapping simplices
(tetrahedra in 3D).
For example, the \emph{Delaunay generalized barycentric coordinates} for a point can be computed by performing a Delaunay tessellation of the polyhedron.
The barycentric coordinates of whichever simplex the point falls inside of
are the generalized barycentric coordinates.
For a 3D point in general position in the interior,
the mixing weights will have at most 4 non-zero weights, which corresponds to
the number of vertices of a tetrahedron.

This is the approach taken by \citet{Tan:2016:DIL} for their \emph{as-sparse-as-possible} (ASAP) technique to extract layers.
\rev{Because \citet{Tan:2016:DIL} considered recursive over compositing,
users provided a layer or vertex order;
they tessellated the simplified convex hull by connecting all its (triangular) faces to the first vertex, which corresponds to the background color. This simple \emph{star tessellation} is valid for any convex polyhedron.
In the additive mixing scenario, no order is provided;
we discuss the choice of tessellation below.}
%, which is akin to computing the Delaunay tessellation of the simplified convex hull, finding which simplex each pixel falls into (or is closest to), and then using the pixel's barycentric coordinates inside the simplex as its mixing weights.
Because the weights are assigned purely based on the pixel's colors,
however,
this approach predictably suffers from spatial coherence artifacts (Figure~\ref{fig:recoloring_different_triangulation}).
% \ref{fig:mixing_weights_map_comparison_between_our_two_methods} \todo{update this figure reference}).
The colors of spatially neighboring pixels may belong to different tetrahedra.
As a result, ASAP layers produce speckling artifacts during operations like recoloring
(Figure~\ref{fig:recoloring_different_triangulation}).
% \yotam{Not ready to say this yet: The layer decomposition must also be recomputed entirely when the palette colors change.}

% \input{star_tessellation}

% \begin{figure*}
% \centering
% \includegraphics[width=\linewidth]{figs/image_decomposition_comparison/results_comparison-2.pdf}
% \caption{\jianchao{should remove} Mixing weights comparison between our proposed RGBXY method and \protect\citet{Tan:2016:DIL} (ASAP
% and the quadratic, additive mixing adaptation of their optimization).
% The quadratic technique occasionally makes undesirable tradeoffs
% between its sparsity and smoothness terms, which manifest as awkward layer boundaries.
% The ASAP algorithm does not maintain spatial smoothness.
% While somewhat less sparse, our proposed 5D RGBXY convex hull-based approach naturally considers spatial and color proximity.}
% \label{fig:mixing_weights_map_comparison_between_our_two_methods}
% \end{figure*}

\paragraph{Spatial Coherence}
To provide spatial coherence,
our key insight is to extend this approach to 5D RGBXY-space,
where XY are the coordinates of a pixel in image space, so that spatial relationship
are considered along with color in a unified way (Figure~\ref{fig:RGBXY_method_illustration}).
We first the compute convex hull of the image in RGBXY-space.
We then compute Delaunay generalized barycentric coordinates (weights) for every pixel in the image in terms of the 5D convex hull.
Pixels that have similar colors \emph{or} are spatially adjacent will end up with similar weights, meaning that our layers will be smooth both in RGB and XY-space.
These mixing weights form an $Q \times N$ matrix $W_\text{RGBXY}$,
where $N$ is the number of image pixels and $Q$ is the number of RGBXY convex hull vertices.
We also compute $W_\text{RGB}$, Delaunay barycentric coordinates (weights) for the RGBXY convex hull vertices in the 3D simplified convex hull.
We use the RGB portion of each RGBXY convex hull vertex, which always lies inside the RGB convex hull.
Due to the aforementioned out-of-gamut projection step when computing the simplified RGB convex hull, however, an RGBXY convex hull vertex may occasionally fall outside it. We set its weights to those of the closest point on the 3D simplified convex hull.
$W_\text{RGB}$ is a $P \times Q$ matrix, where $P$ is the number of vertices of the simplified RGB convex hull (the palette colors).

% Any linear projection of vertices of the convex hull of a set of points after projecting down to a lower dimension are also vertices of the convex hull of the original points.}

The final weights for the image are obtained via matrix multiplication: $W = W_\text{RGB} W_\text{RGBXY}$, which is a $P \times N$ matrix that assigns
each pixel weights solely in terms of the simplified RGB convex hull.
These weights are smooth both in color and image space.
To decompose an image with a different RGB-palette,
one only needs to recompute $W_\text{RGB}$ and then perform matrix multiplication.
Computing $W_\text{RGB}$ is extremely efficient,
since it depends only on the palette size and the number of RGBXY convex hull vertices.
It is independent of the image size and allows users to experiment with
image decompositions based on \rev{interactive palette editing
(Figure~\ref{fig:recoloring_GUI_editing_before_after} and the supplemental materials).}

\begin{figure}
	\centering
	\includegraphics[width=\linewidth]{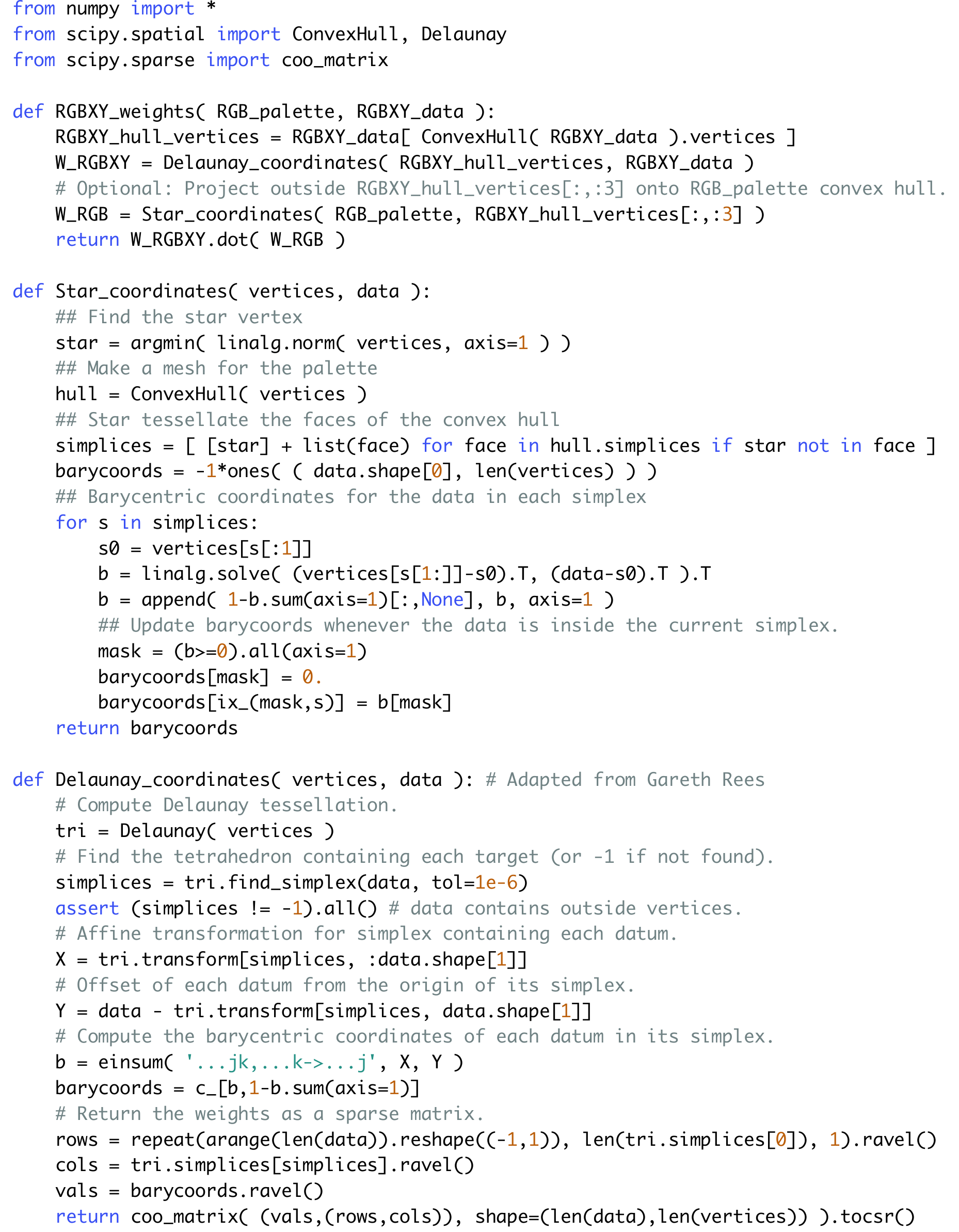}
	\caption{\rev{Python code for our RGBXY additive mixing layer decomposition (48 lines).}}
	\label{fig:decomposition_algorithm}
\end{figure}

\revbegin
\paragraph{Tessellation}
At first glance, any tessellation of 3D RGB-space has approximately the same $\ell_0$ weight sparsity (4 non-zeros). In practice, the ``line of greys'' between black and white is critically important. Any pixel near the line of greys can be expressed as the weighted combination of vertices in a number of ways (e.g.\ any tessellation).
It is perceptually important that the line of greys be 2-sparse
in terms of an approximately black and white color,
and that nearby colors be nearly 2-sparse.
If not, then grey pixels would be represented as mixtures of complementary colors;
any change to the palette that didn't preserve the complementarity relationships
would turn grey pixels colorful (Figure~\ref{fig:recoloring_different_triangulation}).
This tinting is perceptually prominent and undesirable.\footnote{For pathalogical images containing continuous gradients between complementary colors,
this tinting behavior would perhaps be desired.}
\todo{figure showing a pixel near the diagonal and with the line of grey represented}.

To make the line of greys 2-sparse in this way, the tessellation should ensure that
an edge is created between the darkest and lightest color in the palette.
Such an edge is typically among the longest possible edges through the interior of the polyhedron, as the luminance in an image often varies more than chroma $\times$ hue.
As a result, the Delaunay tessellation frequently excludes the most desirable edge
through the line of greys.
We propose to use a star tessellation.
%; which is trivial to implement and guaranteed
%to exist for a convex polyhedron in 3D.
%\jianchao{Our star triangulation method is described in Algorithm~\ref{algo:star_triangulation}.}
%A star tessellation connects a vertex to all triangular faces.
If either a black or white palette color is chosen as the star vertex,
it will form an edge with the other.
We choose the darkest color in the palette as the star vertex.
This strategy is simple and robust
and extends naturally to premultiplied alpha RGBA images.

We also experimented with a variety of strategies to choose the tessellation
such that the resulting layer decomposition is as sparse as possible:
RANSAC line fitting and PCA on the RGB point cloud and finding the longest edge.
We evaluated the decompositions with several sparsity metrics (\cite{Tan:2016:DIL,aksoy2017unmixing,levin2008spectral}, as well as the fraction of pixels with transparency above a threshold).
Ultimately, tinting was more perceptually salient than changes in sparsity, and our
proposed tessellation algorithm is simpler and robust.
\revend

\subsection{Evaluation}
\label{sec:harmony_Palette_Decompostion:comparison}
% \jianchao{Some figures to show two methods' results, their sparsity, their reconstruction RMSE, their running time, their artifacts in layer results and their recoloring comparison. Already done, just lay out the results into figure}

\begin{figure*}
\centering
\includegraphics[width=.475\linewidth]{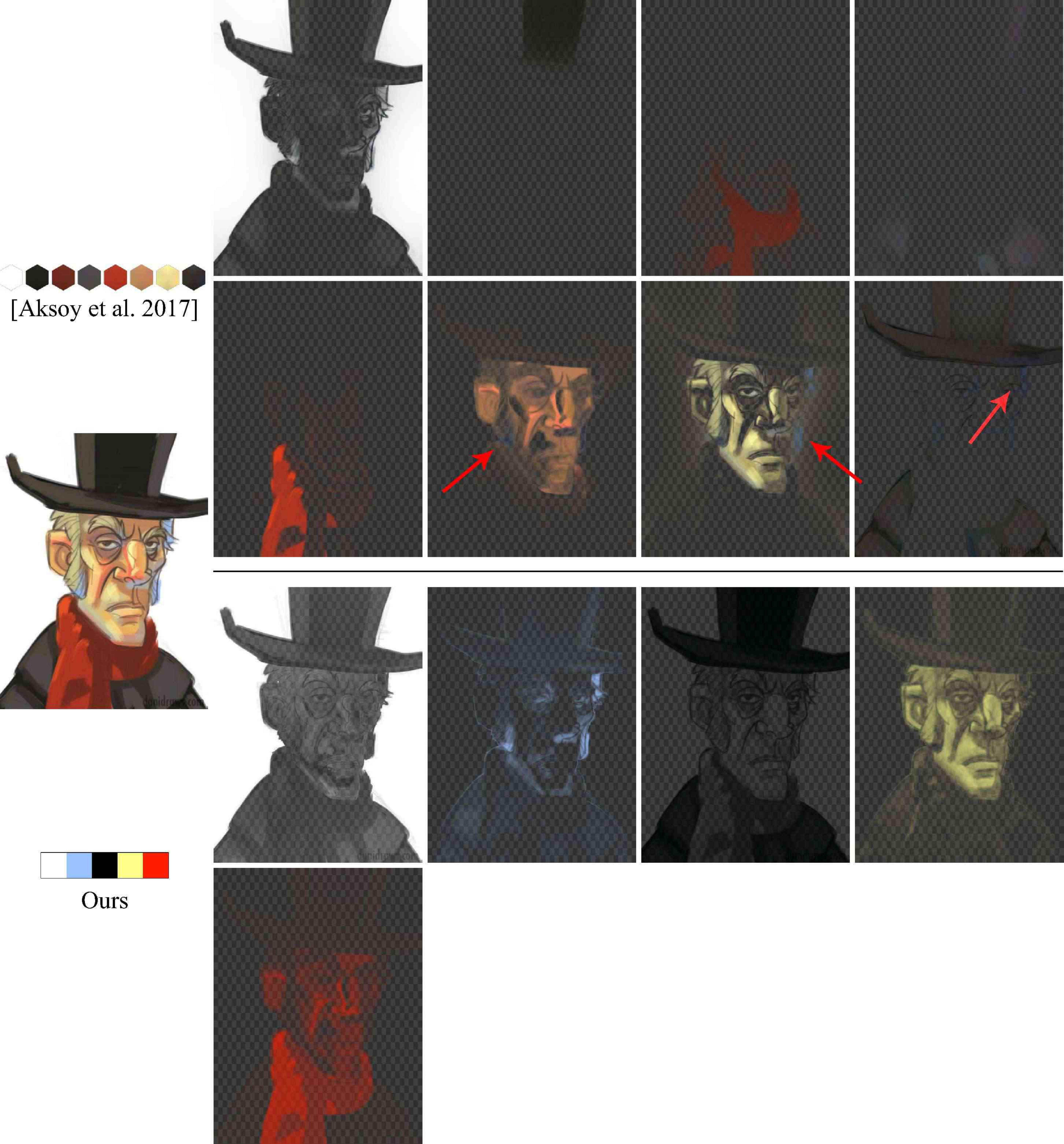}
\includegraphics[width=.51\linewidth]{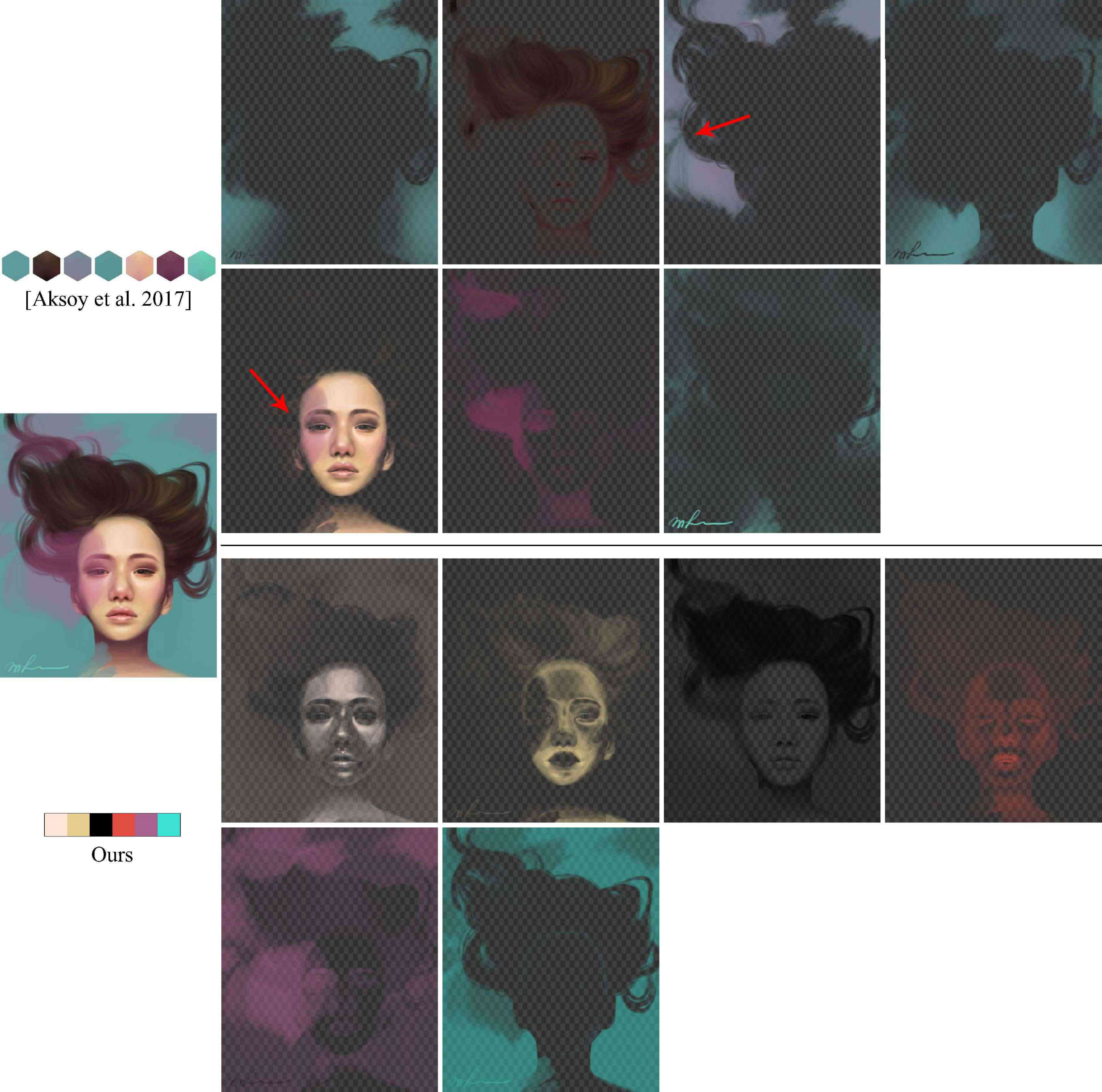}
\caption{A comparison between our proposed RGBXY image decomposition and that of \protect\citet{aksoy2017unmixing}.
\protect\citet{aksoy2017unmixing} creates an overabundance of layers (two red layers above) and does not extract the blueish tint, which appears primarily in mixture.
Our RGBXY technique identifies mixed colors is able to separate the translucent purple haze in front of the girl's face.
\rev{Additionally, our GUI allows editing the palette to modify layers in real time.
This allows users to further improve the decomposition, as shown in
Figure~\ref{fig:recoloring_GUI_editing_before_after} and the supplemental materials.}}
\label{fig:mixing_weights_map_comparison_aksoy}
\end{figure*}

\begin{figure*}
\centering
\includegraphics[width=\textwidth]{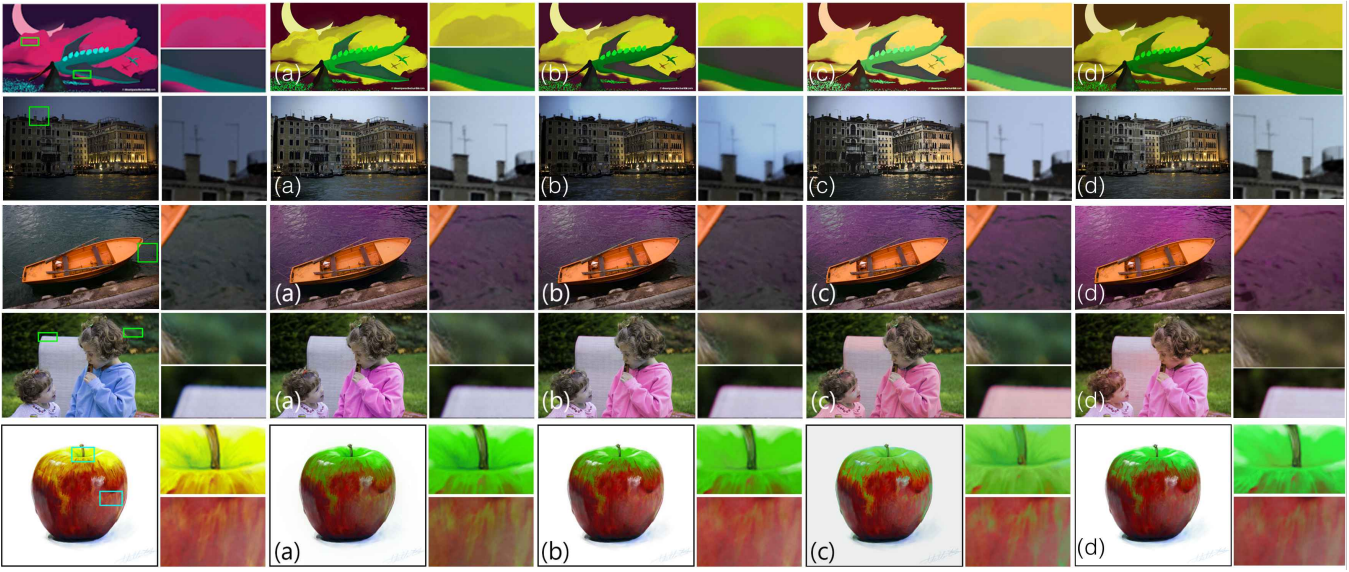}
\caption{\revbegin To evaluate our RGBXY decomposition algorithm,
we compare our layers with previous approaches in a recoloring application.
From left to right: (a) \protect\citet{aksoy2017unmixing}, (b) \protect\citet{Tan:2016:DIL}, (c) \protect\citet{Chang:2015:PPR} and (d) our approach. Our recoloring quality is similar to the state of the art, but our method is orders of magnitude faster and allows interactive layer decomposition while editing palettes.}
\label{fig:recoloring_comparison_all}
\end{figure*}

\begin{figure}
\centering
\includegraphics[width=\columnwidth]{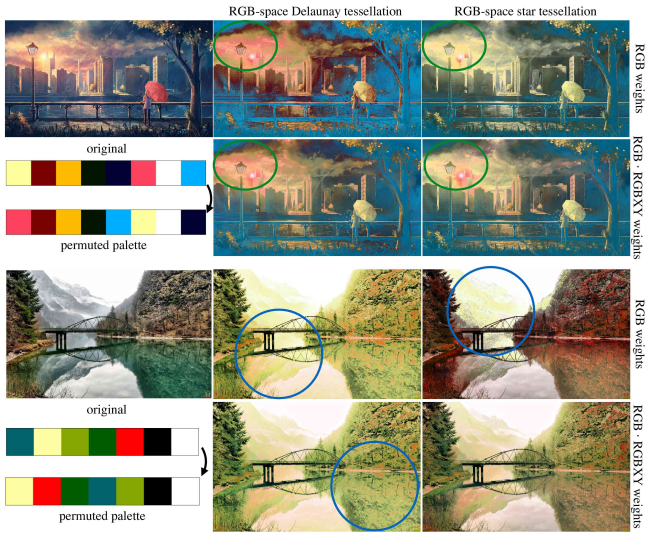}
\caption{\revbegin Comparing tessellation strategies for color palettes in RGB-space.
The Delaunay tessellation column computes Delaunay barycentric coordinates for the
color palette.
This tessellation often avoids creating the perceptually important line of greys, leading to tinting artifacts. These are avoided with a star tessellation emanating from the vertex closest to black.
Computing weights in RGB-space alone leads to spatial smoothness artifacts.
Our two-stage RGBXY decomposition provides color and spatial smoothness.
To interrogate the quality of layer decompositions, we randomly permute the palette,
revealing problems in computed weights.
See the supplemental materials for additional examples.}
\label{fig:recoloring_different_triangulation}
\end{figure}

\begin{figure}
\centering
\includegraphics[width=\columnwidth]{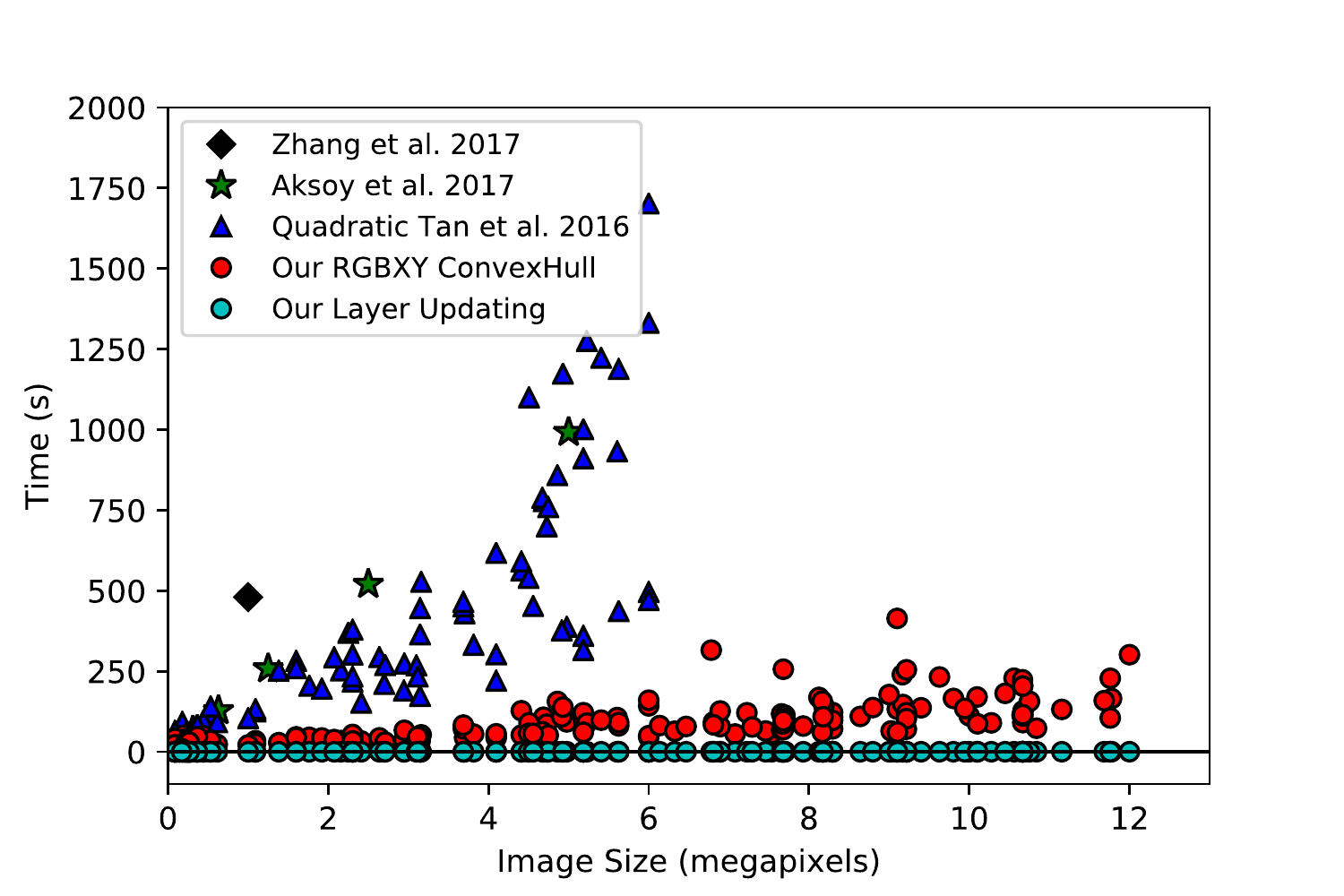}
\caption{
Running time comparison between four additive mixing image decomposition algorithms.
We evaluated our RGBXY algorithm on 170 images up to 12 megapixels
and an additional six 100 megapixel images (not shown; average running time 12.6 minutes).
Our algorithm's performance is orders of magnitude faster
and scales extremely well with image size.
% Occasional outliers (three red dots somewhat above the others)
% illustrate that
The number of RGBXY convex hull vertices has a greater effect on performance than image size.
\rev{Re-computing our layer decomposition with an updated palette is nearly instantaneous
(a few to tens of milliseconds).}
}
\label{fig:four_decomposition_method_time_comparison}
\end{figure}

\revbegin

\paragraph{Quality}
The primary means to assess the quality of layers is to apply them for some purpose,
such as recoloring, and then identify artifacts, such as noise, discontinuities,
or surprisingly affected regions.
Figure~\ref{fig:recoloring_comparison_all} compares recolorings created with
our layers to those from
\citet{aksoy2017unmixing}, \protect\citet{Tan:2016:DIL}, and \citet{Chang:2015:PPR}.
Our approach generates recolorings without discontinuities (the sky in (b), second row),
undesirable changes (the top of the chair in (c), third row), or noise.
% Our approach is somewhat less sparse than.

\revend
We have no explicit guarantees about the sparsity of our weights.
$W_\text{RGB}$ is as sparse as possible to reconstruct 3D colors (4 non-zeros).
$W_\text{RGBXY}$ has 6 non-zeros among the (typically) 2000--5000 RGBXY convex hull vertices, which is also as sparse as possible to recover a point in RGBXY-space.
The sparsity of the product of the two matrices depends on which 3D tetrahedra
the 6 RGBXY convex hull vertices fall into.
\revbegin
Nevertheless, it can be seen that our results' sparsity is almost as good as
\citet{Tan:2016:DIL}.

Figure~\ref{fig:mixing_weights_map_comparison_aksoy} shows a direct
comparison between our additive mixing layers and those of
\citet{aksoy2017unmixing} for direct inspection.
In contrast with our approach, \citet{aksoy2017unmixing}'s approach
has trouble separating colors that appear primarily in mixture.
As a result, \citet{aksoy2017unmixing}'s approach sometimes creates
an overabundance of layers, which makes recoloring tedious,
since multiple layers must be edited.

Our decomposition algorithm is able to reproduce input images
virtually indistinguishably from the originals.
For the 100 images in Figure~\ref{fig:four_decomposition_method_time_comparison},
our RGBXY method's RGB-space RMSE is typically $2--3$.
% The \citet{Tan:2016:DIL} variant's optimization method produces images
% with an average RMSE of 4.1.
% The additional error is due to a smoothness/reconstruction tradeoff made by the objective function.
%
Aksoy et al.~\shortcite{aksoy2017unmixing}'s algorithm reconstruct images with zero error, since their palettes are color distributions rather than fixed colors.

We evaluate our RGB tessellation in Figure~\ref{fig:recoloring_different_triangulation}.
In this experiment, we generate a random recoloring by permuting the colors in the palette.
The RGB-space star triangulation approach is akin to \citet{Tan:2016:DIL}'s
ASAP approach with the black color chosen to be the first layer.
The lack of spacial smoothness is apparent in between the RGB-only decompositions
in the top-row and the RGBXY decompositions in the bottom row.
The decompositions using the Delaunay generalized barycentric coordinates (left column)
result in undesirable tinting for colors near the line of grey.
Additional examples can be found in the supplemental materials.

Throughout the remainder of the paper,
all our results rely on our proposed layer decomposition.

% \citet{Tan:2016:DIL} occasionally makes undesirable tradeoffs
% between its sparsity and smoothness terms, which manifest as awkward layer boundaries.
% Their ASAP algorithm does not maintain spatial smoothness.
% While somewhat less sparse, our proposed 5D RGBXY convex hull-based approach naturally considers spatial and color proximity.

\revend

\paragraph{Speed}
In Figure~\ref{fig:four_decomposition_method_time_comparison}, we compare the running time
of additive mixing layer decomposition techniques.
We ran our proposed RGBXY approach on 100 images under 6 megapixels
with an average palette size of 6.95 and median palette size of 7.
Computation time for our approaches includes palette selection (RGB convex hull simplification).
Because of its scalability, we also ran our proposed RGBXY approach on an additional 70 large images between 6 and 12 megapixels,
and an additional 6 extremely large images containing 100 megapixels (not shown in the plot).
The 100 megapixel images took on average 12.6 minutes to compute.
Peak memory usage was 15 GB.
For further improvement, our approach could be parallelized
by dividing the image into tiles, since the convex hull of a set of convex hulls
is the same as the convex hull of the underlying data.
A working implementation of the RGBXY decomposition method can be found in Figure~\ref{fig:decomposition_algorithm} (48 lines of code).
\rev{The ``Layer Updating'' performance is nearly instantaneous, taking a few milliseconds to, for 10 MP images, a few tens of milliseconds to re-compute the layer decomposition given a new palette.}

Our running times were generated on a 2015 13'' MacBook Pro with a 2.9 GHz Intel Core i5-5257U CPU and 16 GB of RAM.
Our layer decomposition approach was written
in non-parallelized Python using NumPy/SciPy and their wrapper for the QHull convex hull and Delaunay tessellation library \cite{QHull}.
\rev{Our layer updating was written in OpenCL.}

% \jianchao{can we say our RGBXY time complexity are sublinear complexity? }
% \yotam{I don't think we can make an explicit O(n) claim.
% It might be linear with a very small slope. It should be whatever convex hull is!}
%
\citet{aksoy2017unmixing}'s performance is the fastest previous work known to us.
The performance data for Aksoy et al.'s algorithm is as reported in their paper
and appears to scale linearly in the pixel size.
Their algorithm was implemented in parallelized C++.
\citet{aksoy2017unmixing} reported that their approach took 4 hours and 25 GB of memory to decompose a 100 megapixel image.
\citet{zhang2017palette}'s sole performance data point is also as reported in their paper.

We also compare our approach to a variant of \citet{Tan:2016:DIL}'s optimization.
We modified their reconstruction term to the simpler, quadratic one that matches our additive mixing layer decomposition scenario. With that modification, all energy terms become quadratic. However, because the sparsity term is not positive definite, it is not a straightforward Quadratic Programming problem; we minimize it with \textsc{L-BFGS-B} and increased the solver's default termination thresholds since RGB colors have low precision (gradient and function tolerance $10^{-4}$).
This approach was also written in Python using NumPy/SciPy.
The performance of the modified \citet{Tan:2016:DIL} is somewhat unpredictable,
perhaps owing to the varying palette sizes.

The fast performance of our approach is due to the fact that the number of RGBXY convex hull vertices $Q$ is virtually independent of the image size
and entirely independent of the palette size.
Finding the simplex that contains a point is extremely efficient (a matrix multiply followed by a sign check) and scales well.
Our algorithm's performance is more correlated with the number of RGBXY convex hull vertices and tessellated simplices.
\rev{This explains the three red dots somewhat above the others in the performance plot.}
% Our technique makes use of
% optimized, commonly used routines from computational geometry~\cite{QHull}.
% The performance of optimization-based method depends
% more directly on the number of pixels.
% Numerical and visual comparison of two methods are shown below.
% is a simple but elegant geometry method for image decomposition problem, its implementation code is very neat, we put the code in appendix. \todo{put code in appendix? or put three lines pseudo code here?}
% \yotam{Great idea! Put it here!}

In contrast, optimization-based approaches typically have parameters to tune,
\rev{such as the balance between terms in the objective function,
iteration step size, and termination parameters.}

\begin{figure}
\centering
\includegraphics[width=\linewidth]{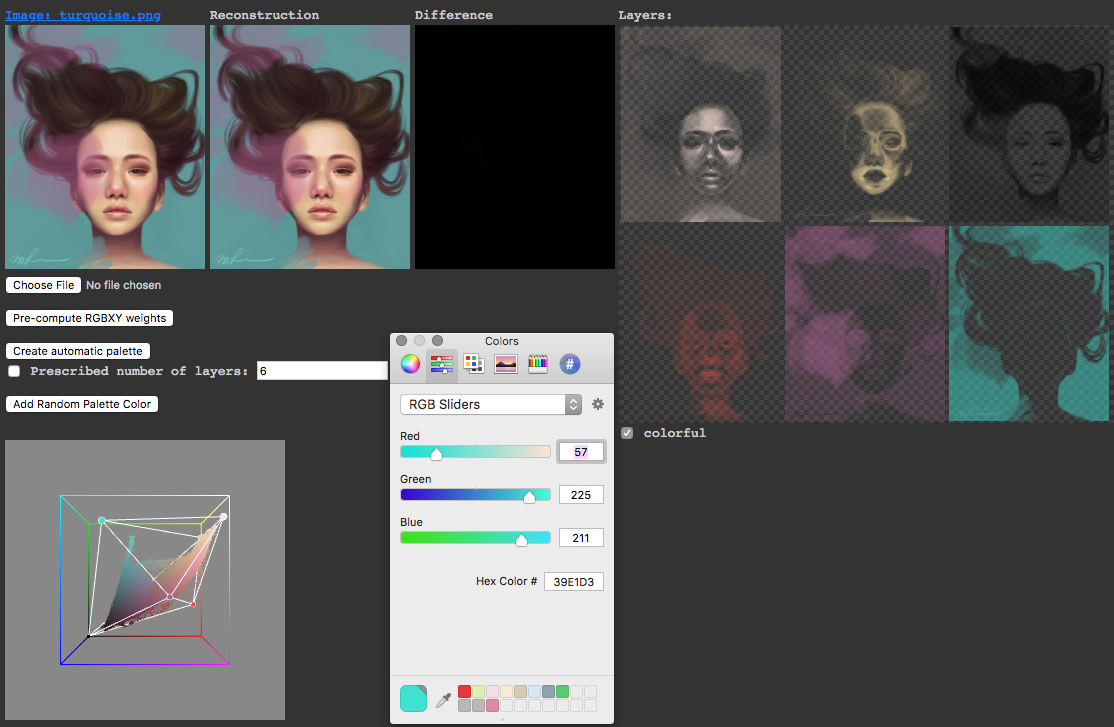}
\caption{\rev{Our GUI for interactively editing palettes. See the text for details.}}
\label{fig:palette_editing_gui}
\end{figure}

\revbegin
\paragraph{Interactive Layer Decompositions}
To take advantage of our extremely fast layer decomposition,
we implemented an HTML/JavaScript GUI for viewing and interacting with layer decompositions (Figure~\ref{fig:palette_editing_gui}).
An editing session begins when a user loads an image and precomputes RGBXY weights.
Users can then begin with a generic tetrahedron or with an automatically chosen palette,
optionally with a prescribed number of layers.
Users can alter the palette colors in 3D RGB-space (lower-left)
or activate a traditional color picker by clicking on a layer (the turquoise layer as shown).
As users drag the palette colors, the layer decomposition updates live.
(Although our layer updating algorithm computes at an extremely high frame rate,
the bulk of the time in our GUI is spent transferring the data to the browser via a WebSocket.)
Users can also add and then manipulate additional colors.
See Figure~\ref{fig:recoloring_GUI_editing_before_after} for a result created with our GUI; see the supplemental materials for screen recordings of this and other examples.
\revend

\begin{figure}
\centering
\includegraphics[width=\linewidth]{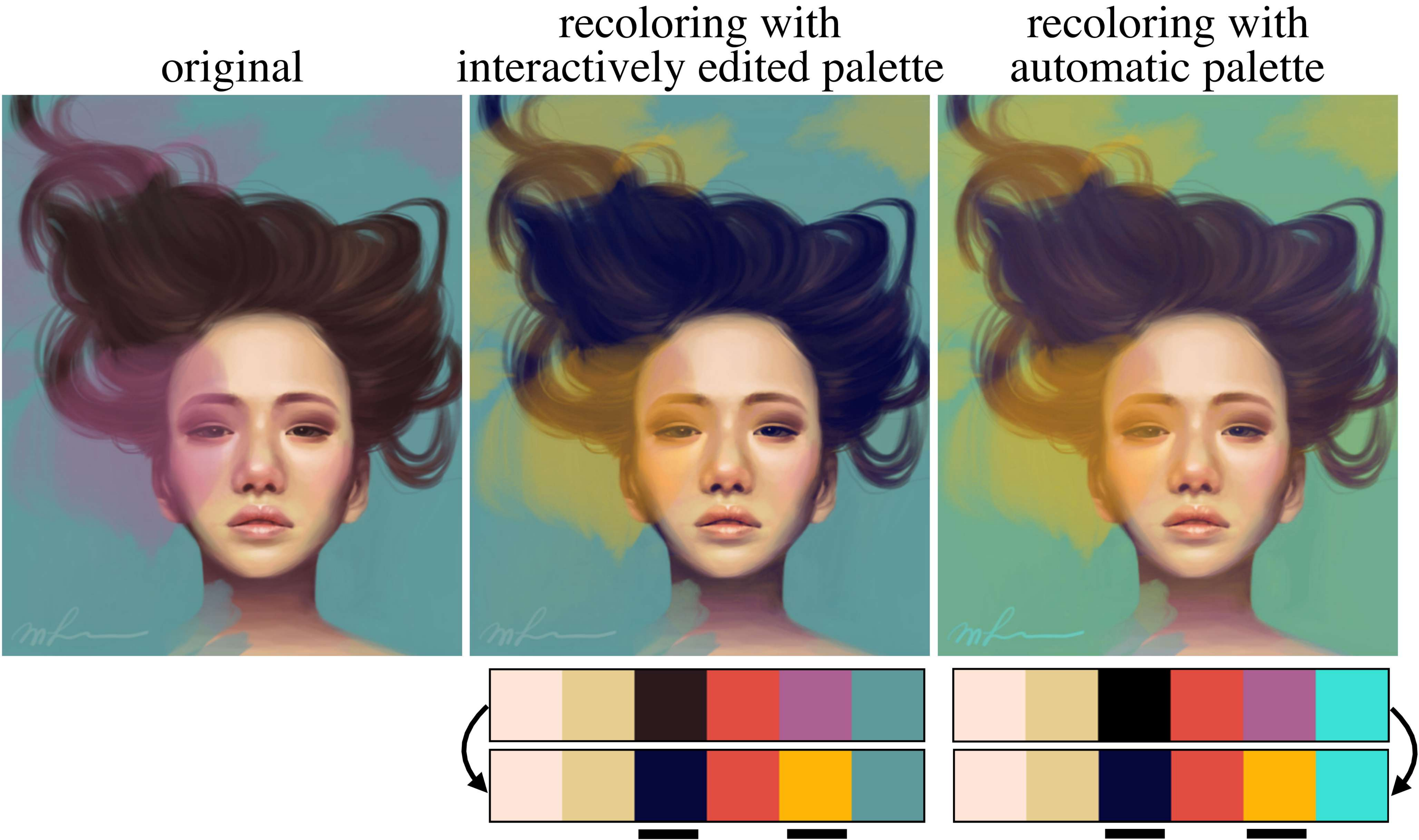}
\caption{\revbegin Our GUI allows users edit palettes and see the resulting
layer decomposition in real-time.
Videos of live palette editing can be found in the supplemental materials.
In this example, the automatic palette (right) becomes sparser
as a result of interactive editing. The user edits the automatically generated
palette to ensure that the background and hair colors are directly represented.
As a result, editing the purple haze and hair no longer affects the background color.}
\label{fig:recoloring_GUI_editing_before_after}
\end{figure}

%% file: harmonization.tex
\section{Color Harmonization}
\label{sec:harmony_Harmonization}

In the following we describe our palette-based approach to color harmonization and color composition. Our work is inspired by the same concepts and goals as related previous work~\cite{Cohen-Or:2006:CH:1179352.1141933}. However, we also aim for a simpler and more compact representation
that can express additional operations and be applied directly to palettes.
First, we explain how we fit and enforce classical harmonic templates. Next, we describe how our framework can be used for other color composition operations. 

\begin{figure}
	\centering
	\includegraphics[width=\columnwidth]{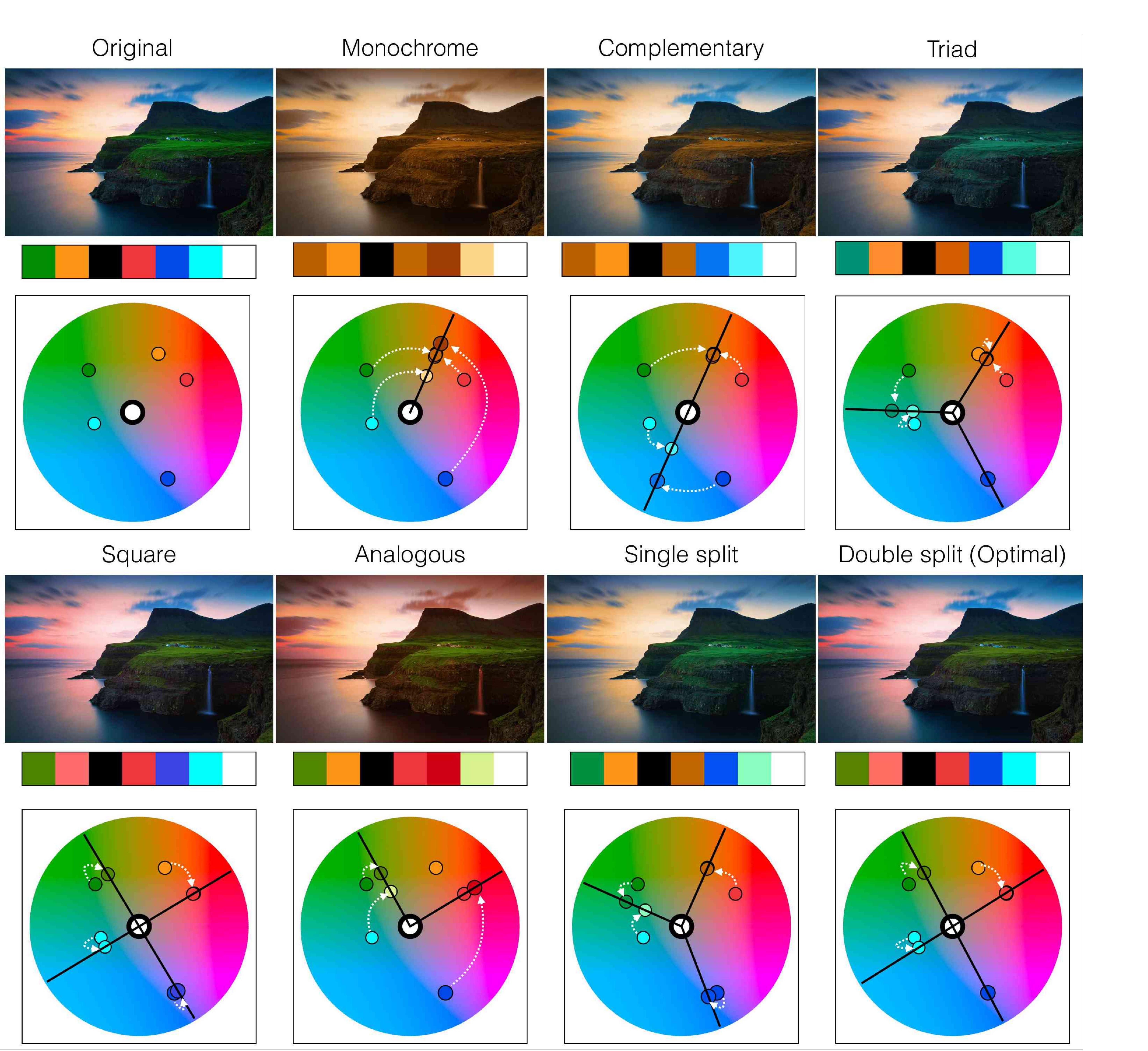}
	\caption{Results of our harmonic templates fit to the an input image. We can see how each of them is able to provide a balanced and pleasing look when the harmonization is fully enforced.
	In an interactive application, the user can control the strength of harmonization,
	which interpolates the hue rotation of each palette color.}
	% parameter  additional global rotation $\beta$ over the fitted $T_m(\alpha^*)$.}
	\label{fig:color_harmonization-full}
\end{figure}

\subsection{Template fitting}
\label{sec:harmony_Harmonization: template fitting}
%\citet{Cohen-Or:2006:CH:1179352.1141933} defined some harmonization templates on HS wheels, where H is Hue and S is Saturation. Their templates consist of different size sectors, as showed in Figure~\ref{fig:template_axis}. All pixel colors in HS wheels are mapped into sectors to enable final image color harmonization. Our templates consist of several axises, as showed in Figure~\ref{fig:template_axis}. And we choose to use LCH color space (Luminance, Chroma, Hue) to do color harmonization, since in HSV space, changing Hue value will affect Luminance, which may cause some unnatural image recoloring results; however, in LCH space, Luminance channel is independent from Hue channel, so the modification of Hue value will not affect image Luminance at all. 

%As showed in Figure~\ref{fig:template_axis}, we have seven templates: $T_m, m=1 ... 7$. Given an image $I$ and its extracted palette colors $P$, we will find the template $T_m$ that is closest to palette colors $P$ in HC wheel, and we will project each palette color onto their closest corresponding template axis to obtain harmonized palette colors, then mix modified palette using our precomputed mixing weights to obtain final harmonized image. The template is defined by $T(m, \alpha)$, where $m$ is index of template, and $\alpha$ is rotation angles of template in HC wheel. Specifically. We define a distance between a palette $P$ and a template $T_m$ in HC wheel as equation~\ref{eq:harmony:palette_tempalte_distance} to measure closeness of template and palette.

Figure~\ref{fig:template_axis} shows our new axis-based templates compared to the sector-based ones from~\citet{tokumaru2002color}. For our results in this paper we use seven templates $T_m, m=1 ... 7$. A template is defined by $T_{m}^j(\alpha)$, where $j$ is the index of each axis (the total number of axes varies between templates), and $\alpha$ is an angle of rotation in hue. While our templates are valid in any circular (or cylindrical) color space (e.g. HSV), we apply them in LCh-space (Lightness, Chroma, and hue) to match human perception.

Given an image $I$ and its extracted color palette $P$, we seek to find the $T_m(\alpha)$ that is closest to the colors in $P$ in the Ch plane. For that, we find the closest axis to each color, and solve for the global rotation and additional angles that define the template.  We define the distance $D$ between a palette $P$ and a template $T_m(\alpha)$ as:
%equation~\ref{eq:harmony:palette_tempalte_distance}:
%
\begin{align}
%\begin{split}
\label{eq:harmony:palette_tempalte_distance}
D(P, T_{m}(\alpha)) &= \sum_{i=1}^{|P|}{W(P_i) \cdot L(P_i) \cdot C(P_i) \cdot \left| H(P_i)-T_{m}^{j^*}(\alpha) \right| } \\
j^* &= \argmin_{j=1 \hdots \#\text{axes}} \left| H(P_i)-T_{m}^j(\alpha) \right|  \nonumber
%\end{split}
\end{align}
where $j^*$ is the axis of template $T_{m}(\alpha)$ that is closest to palette color $P_i$. $\left| \cdot \right|$ measures the difference in Hue angle. Note that for the \emph{analogous} template, any palette color inside that arc area will be zero distance to the template.
$W(P_i)$ is the contribution of color $P_i$ to all the pixels in image, computed as the sum of all the weights for layer $i$ and normalized by the total number of pixels in the image. $W(P_i)$ promotes the template to be better aligned with the relevant colors of the image. When using color palettes that do not come from images, $W(P_i)$ is the same for each color and can be discarded.
The lightness $L(P_i)$ and chroma $C(P_i)$ of the color are also used as weights so that we measure the arc distance around the color wheel (the angular change scaled by radius). The darker the color or the less saturated, the smaller the perceived change per hue degree.

% solve Equation~\ref{eq:harmony:optimal_angle} to 
Since the search space is a finite range in 1D, we use a brute-force search to find the optimal global rotation angle $\alpha_m^*$ fitting a template $T_m(\alpha)$ to a palette $P$:

\begin{equation}
\label{eq:harmony:optimal_angle}
\alpha_m^* = \argmin_{\alpha} D(P, T_{m}(\alpha))
\end{equation}

\emph{Monochrome}, \emph{complementary}, \emph{triad} and \emph{square} templates have only one degree of freedom, so we search the global rotation every 1 degree in $[0,360]$. 
For \emph{analogous}, \emph{single split} and \emph{double split} we allow an additional degree of freedom (angle between axes), which we allow $[-15, 15]$ degrees. In this case, $\alpha_m^* = [\alpha_{m,1}^*, \alpha_{m,2}^*]$. With $\alpha_{m,1}^*$ being the optimal global rotation, and $\alpha_{m,2}^*$ the optimal angle between axes. Given that palettes are typically small (less than 10 colors), our brute force search is very fast (less than a second).
% around $\|P\|=7$

\begin{figure}
	\centering
	\includegraphics[width=\columnwidth]{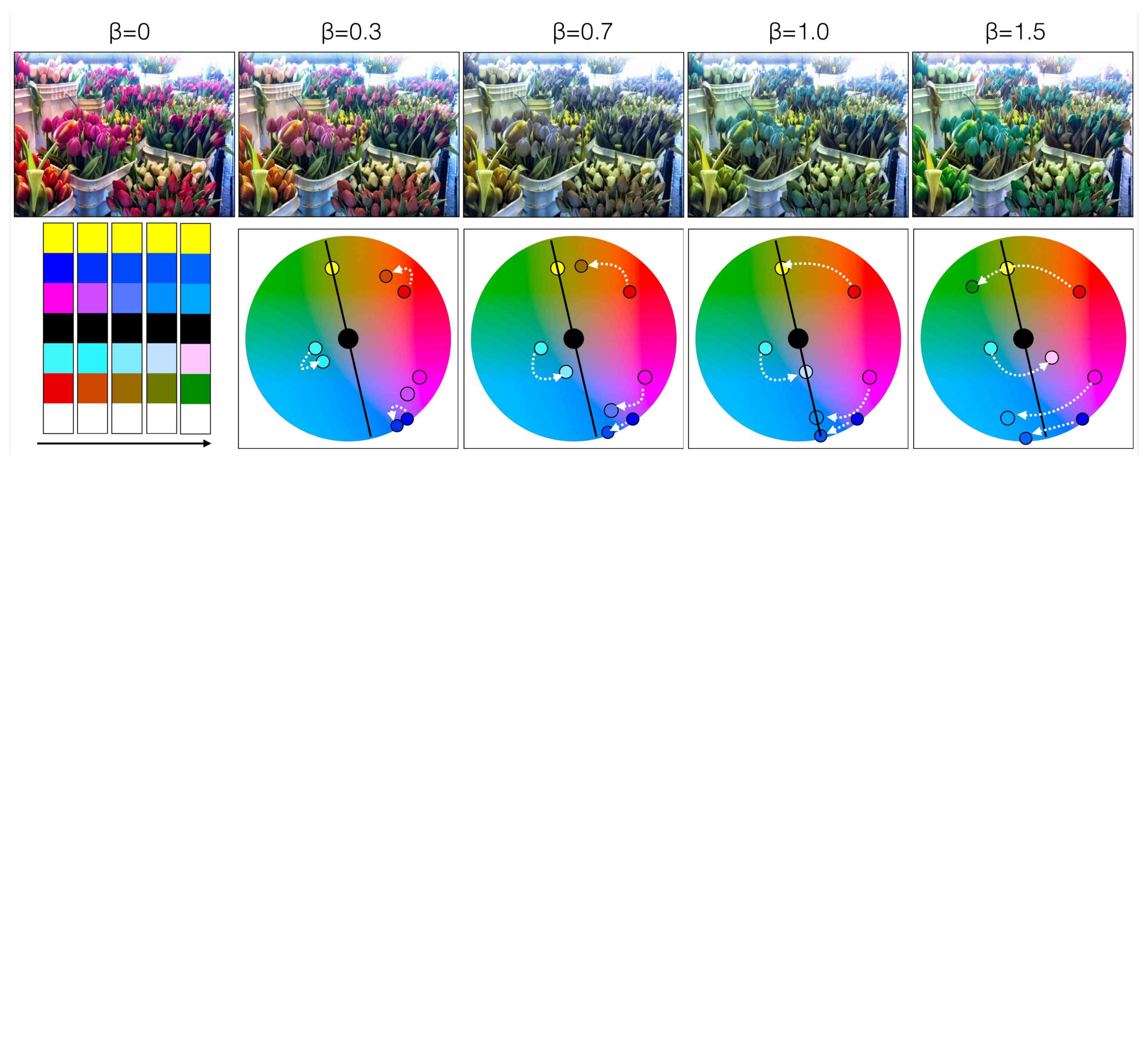}
	\caption{Results from enforcing a given template $T_m(\alpha^*)$ with varying degrees of strength $\beta$. Bottom left shows the consistent palette interpolation across $\beta=[0,1.5]$. Even beyond $\beta=1$ (full harmonization), results remain predictable.}
	\label{fig:color_harmonization-interp}
\end{figure}

Once a template is fit, we harmonize the input image by using $T_m(\alpha_m^*)$ to move the colors in $P$ closer to the axis assignment that minimizes equation~\ref{eq:harmony:palette_tempalte_distance}.
We leverage the image decomposition to recolor the image.
Because we use a spatially coherent image decomposition, no additional work is needed to prevent discontinuous recoloring as in \citet{Cohen-Or:2006:CH:1179352.1141933}.
Figure~\ref{fig:color_harmonization-full} shows different harmonic templates enforced over the same input image. Additional examples can be found in the supplementary material.
% \jose{Do we need an equation here to describe the snapping to axes and the control over the strength of the harmonization? We will need a figure and we may even show negative strengths (disharmonization?).}
%\yotam{Is the following OK?}
Users can control the strength of harmonization via an interpolation parameter, where $\beta=0$ leaves the palette unchanged and $\beta=1$ fully rotates each palette color to lie on its matched axis (Figure~\ref{fig:color_harmonization-interp}). In the LCh color space, this affects hue alone.
%\todo{Jianchao: please put an equation here describing the snapping to axes with a parameter controlling the strength (t for interpolation?)}

Depending on the colors in $P$, some templates are a better fit than others
as measured by Equation~\ref{eq:harmony:palette_tempalte_distance}.
We can determine the optimal template $T_m^*$ automatically:
\begin{equation}
\label{eq:harmony:optimal_template}
T_m^* = \argmin_{T_m} D(P, T_{m}(\alpha_m^*))
\end{equation}
Depending on the palette size or its distribution, some axes may end up without any color assigned to them. We deem those cases not compliant with the intended balance of the harmonic template and remove them from this automatic selection.

\begin{figure}
	\centering
	\includegraphics[width=\columnwidth]{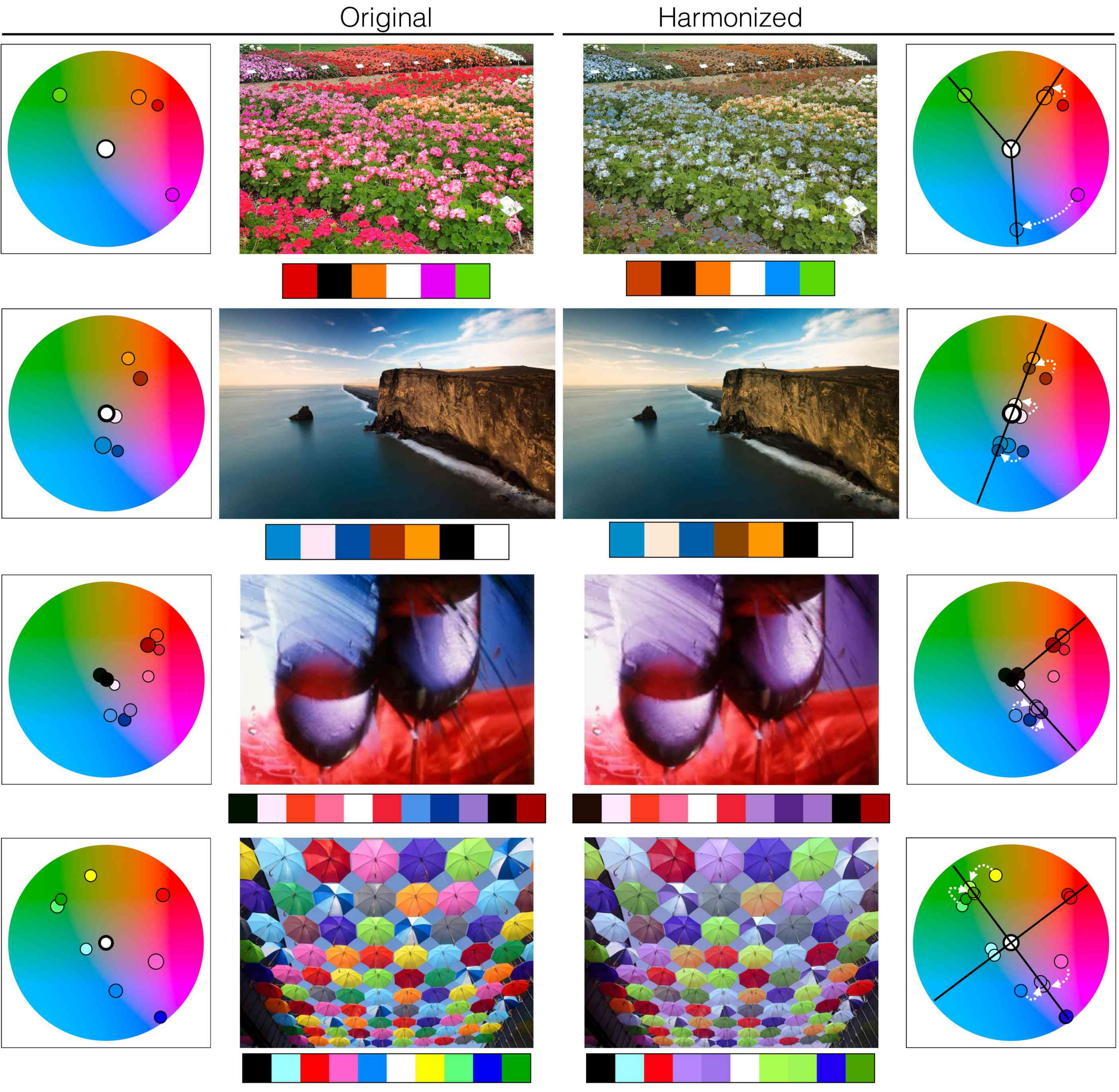}
	\caption{Examples of optimal templates for different images, and the fully harmonized results they produce.}
	%\yotam{This next sentence is probably irrelevant now that we chose examples with bigger changes.} Given that the best fitting template tries to achieve harmony with the minimal change in colors, some of the results may present subtle changes.}
	\label{fig:color_harmonization-more}
\end{figure}

\begin{figure}
	\centering
	\includegraphics[width=\columnwidth]{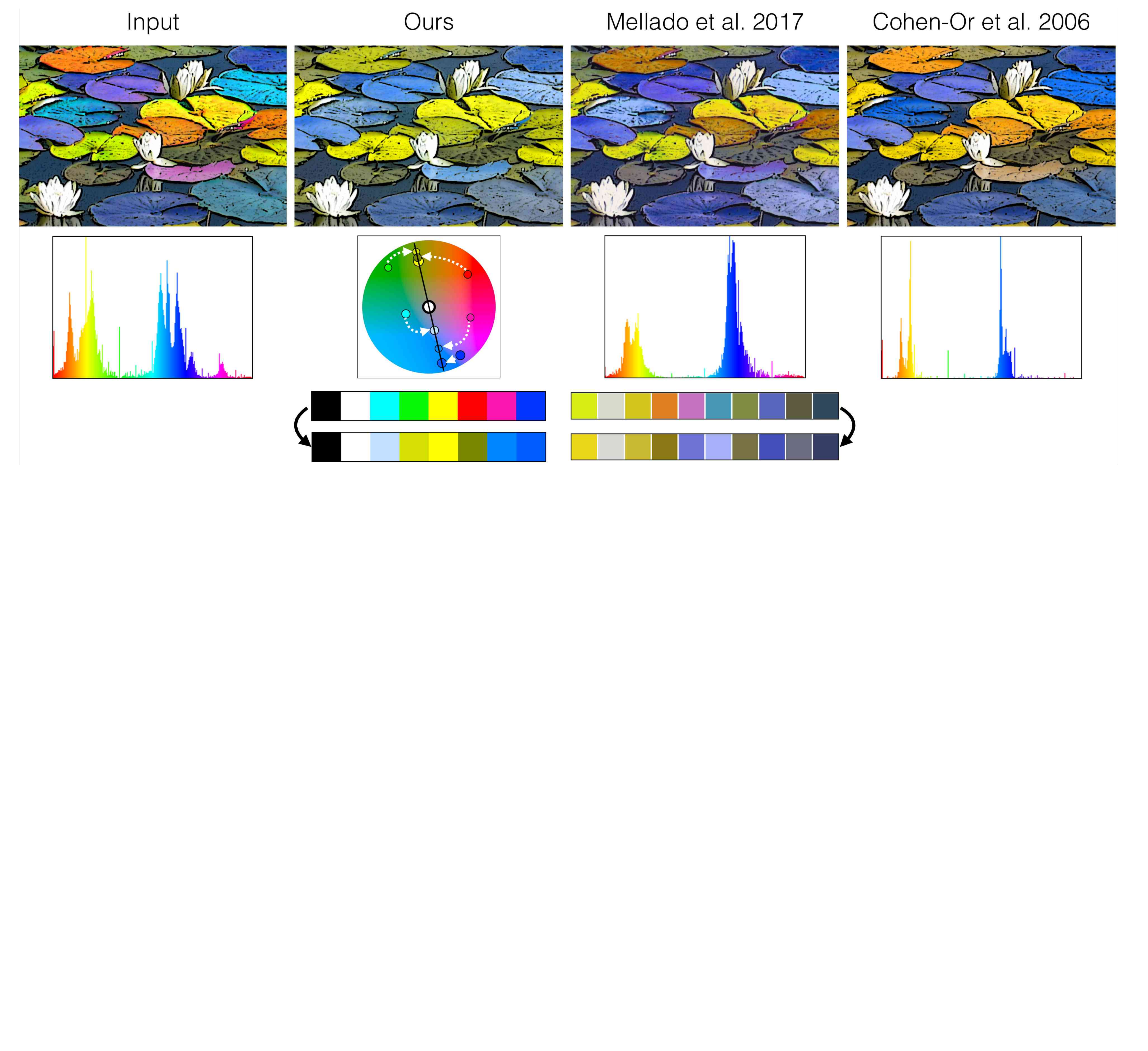}
	\caption{
	%\jianchao{updated.}
	Comparison of harmonizations using the best fitting template from different methods. \citet{Cohen-Or:2006:CH:1179352.1141933} fit a \emph{complementary (I type)} template in HSV space, producing unexpected changes in lightness for some colors. \citet{mellado2017constrained} formulated the same harmonic template using their framework, again in HSV but including additional constraints to preserve lightness.
	Our rotations in LCh-space directly preserve lightness.
	For comparison, we also show our harmonization with a complementary template.
	Our result in this example looks different but \rev{was perceived as more harmonic in a perceptual study ($\chi^2 = 6, p = 0.01$)}.
	Our optimal template according to $T_m^*$ is a \emph{single split}, \rev{which was perceived as similarly harmonic to Cohen-Or et al.'s result}.
	This image harmonized via our other templates can be found in the supplementary material.}
	%\yotam{Keep the following sentence?}
	%None of them look similar to these, but all of them look more \emph{harmonic}.
	%\caption{Our optimal template is single split complementary in LCh space, Cohen-or et al.~\shortcite{Cohen-Or:2006:CH:1179352.1141933} optimal template is complementary, but in HSV space, so the luminance of image will change. Mellado et al.~\shortcite{mellado2017constrained} tried to match Cohen-Or et al.~\shortcite{Cohen-Or:2006:CH:1179352.1141933} results using palette harmonization in HSV space and combine with inter-palette luminance constraints. Our harmonization always keep Luminance unchanged in LCh space.}
	\label{fig:color_harmonization-comparison}
\end{figure}

\begin{figure}
	\centering
	\includegraphics[width=\columnwidth]{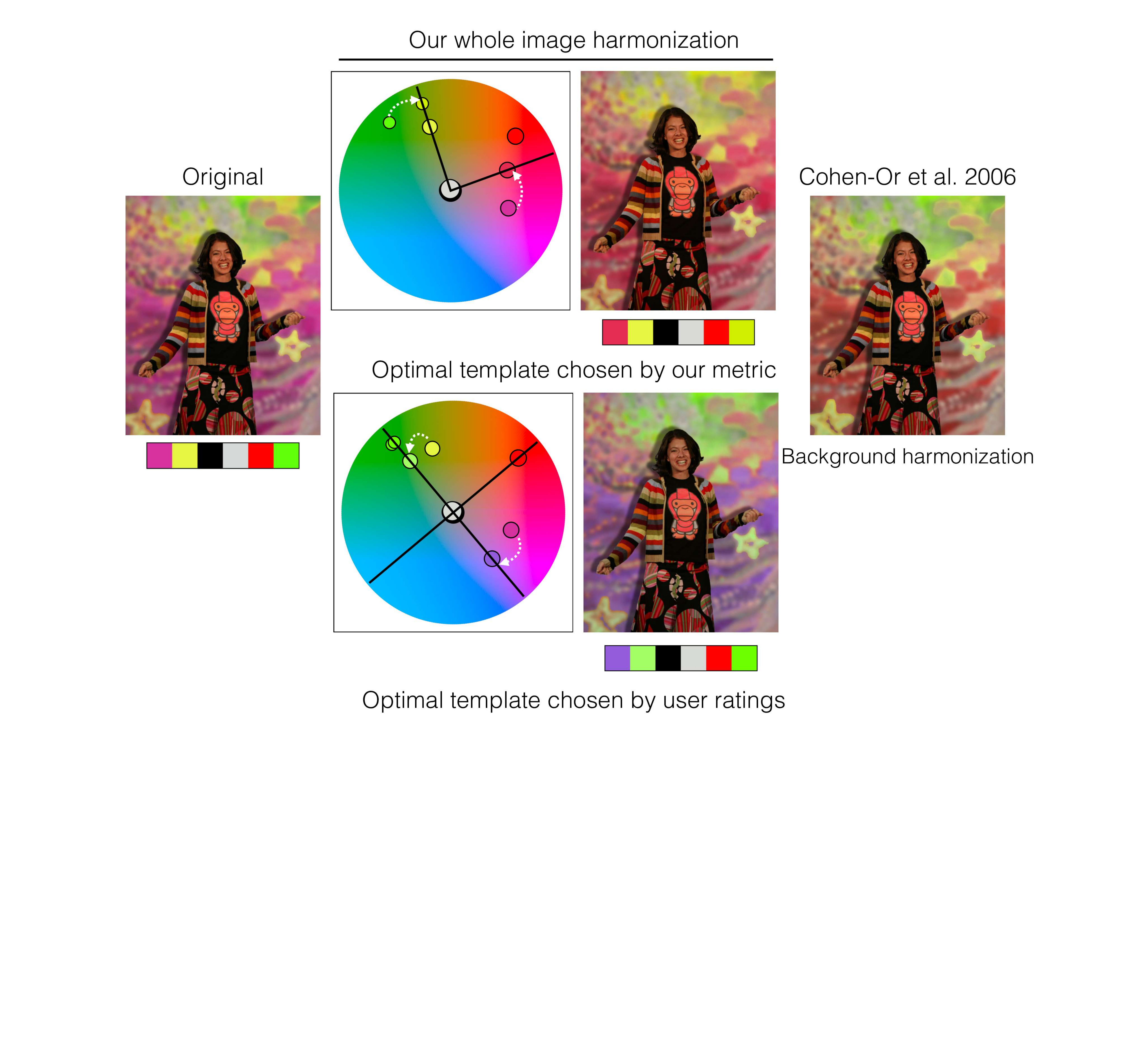}
	\caption{Comparison with masked results from~\citet{Cohen-Or:2006:CH:1179352.1141933}.
	In the top row, Cohen-Or et al.\ harmonized the background to match the colors
	of the masked foreground person.
	We achieve comparable results without masking, and better preserve the background's luminance. With the optimal template (Equation~\ref{eq:harmony:optimal_template}), the harmonized image received similar scores in our perceptual study to the result shown in \citet{Cohen-Or:2006:CH:1179352.1141933}. Below we show the harmonization which received the highest score in the perceptual study (higher than \citet{Cohen-Or:2006:CH:1179352.1141933}'s result).}
	%\yotam{statistically significant difference?}}
	\label{fig:color_harmonization-fg_bg}
\end{figure}

Figure~\ref{fig:color_harmonization-more} shows the best fitting template for a set of images, and the fully harmonized result.
More examples can be found in the supplementary material.
We compare our results with harmonizations from previous works in Figure~\ref{fig:color_harmonization-comparison}. While our result is clearly different, it arguably produces a more balanced result.
%, even when the \emph{single split} template is aiming for contrast.
\citet{Cohen-Or:2006:CH:1179352.1141933}
%and others \jose{Tang?}
demonstrated harmonization between different parts of an image using masks or harmonization of image composites. We provide comparisons for this scenario in Figure~\ref{fig:color_harmonization-fg_bg}.

%The results in Cohen-Or et al.~\shortcite{Cohen-Or:2006:CH:1179352.1141933} shows that their method will cause similar pixel colors mapping to different template sectors, which will cause artifacts in final harmonized results. They use a graph cut optimization to post processing these pixels. Our method does not need this step. Since our palette size is small and representative, no such conflicts will happen. Cohen-Or et al.~\shortcite{Cohen-Or:2006:CH:1179352.1141933} need also post processing step for per pixel hue shift after correctly mapping all pixels to corresponding sectors. Our case does not need this post processing step, because we have small size palette colors and we already pre-compute smooth mixing weights maps, just re-compositing our harmonized palette color using these spatially smooth weights is enough to get final spatially smooth harmonized image, as shown in Figure~\ref{fig:color_harmonization-full} and Figure~\ref{fig:color_harmonization-more}.

%%%this does not make sense now.
% \jianchao{Generalized template fitting for harmonization and color transfer: 7 templates are 7 discrete point in continuous templates space. Thus, we can first fitting from one axis to 4 axis (each axis has its own freedom) for given input palette distribution.  Then find best bipartite matching with one of these 7 predefined templates, so change to that closet template to enable harmonization or do color transferring directly using these axis?? }
% \jianchao{a figure to show these 7 template and their continuous interpolation}

\subsection{Beyond hue}
\label{sec:harmony_Harmonization:beyondHue}

%Compared with sector-based templates, our axes are straightforward to use for other constraints in LCh.
Our compact representation using palettes and axis-based templates allows to formulate full 3D color harmonization operations easily.

\paragraph{LC harmonization}
\label{sec:harmony_Harmonization:LC}

\begin{figure}
	\centering
	\includegraphics[width=\columnwidth]{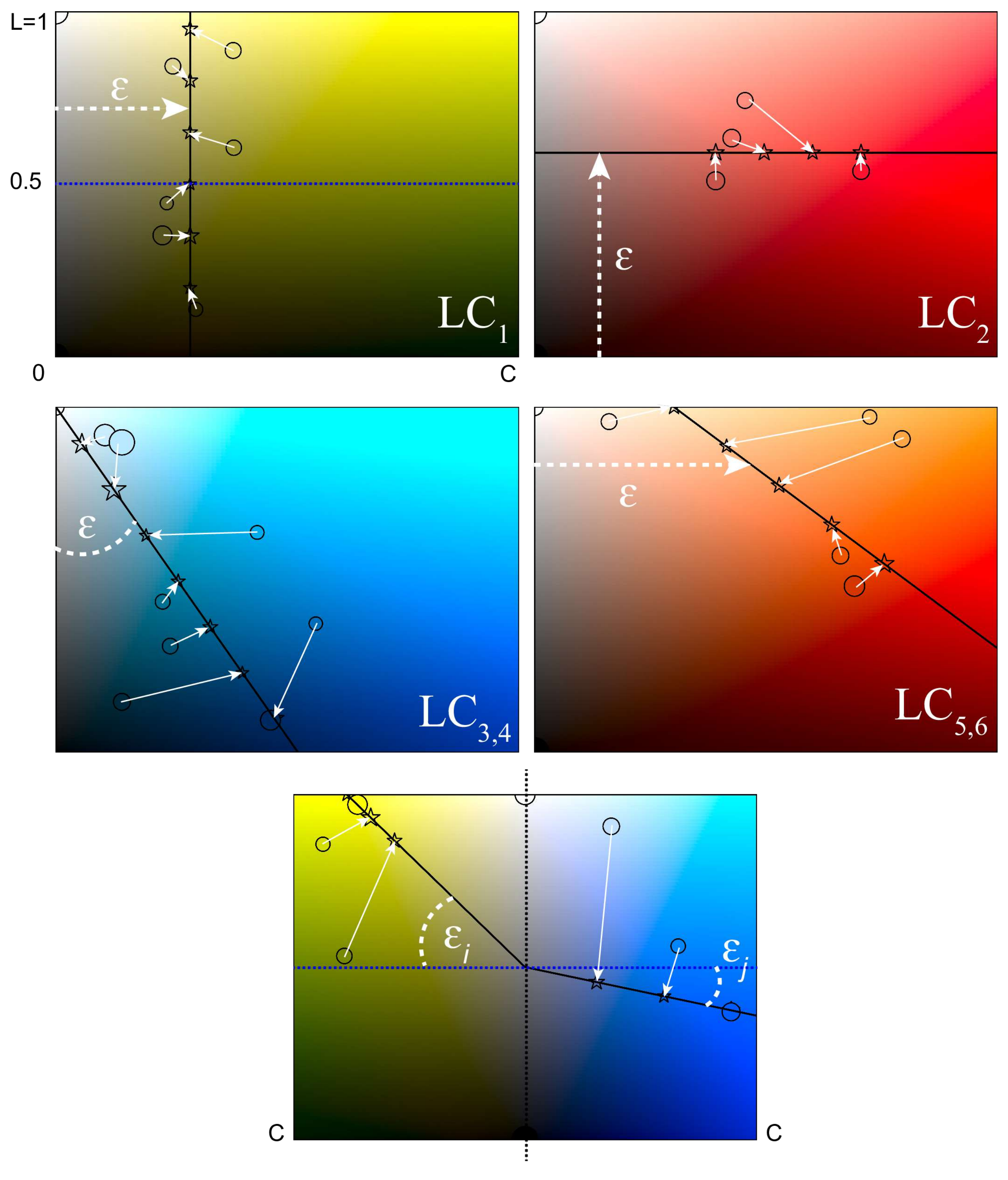}
	\caption{\rev{Our LC templates derived from classical color theory. Template fitting solves for $\epsilon_n$ for each case. From left to right, top to bottom: $LC_1$ aligns colors vertically so hue and chroma remain constant, and places them separated by uniform lightness steps. $LC_2$ enforces constant hue and lightness, and does the uniform distribution for chroma. $LC_3$ and $LC_4$ are alternative diagonal distributions that pivot around $0_{LC}= \{0,0\}$ and $1_{LC}= \{1,0\}$ respectively. $LC_5$ fits colors to a diagonal with an angle of $45^{\circ}$ that displaces horizontally. $LC_6$ is the mirrored version of $LC_5$. For complementary and multi-axis hue templates, $LC_3$ pivots around $N_{LC}=\{0.5,0\}$ for each axis. Additional optional constrains specified in~\cite{Birren1969}, enforce one of the colors to have a neutral lightness of $0.5$. We implement this as a global offset to make the closest color to $L=0.5$ snap into it (blue dotted line), as seen in template $LC_1$. When $P$ includes pure white or black, we found leaving them out of the harmonization works best. Circles show original colors, stars indicate their harmonized location}.  
	}
	\label{fig:LC_templates}
\end{figure}

\rev{Apart from hue, some authors have described harmony in terms of lightness and chroma as well~\cite{Moon44,Birren1969,tokumaru2002color}. While histogram-based approaches may be non-trivial to extend to these additional dimensions, our approach generalizes to them easily. Figure~\ref{fig:LC_templates} shows our interpretation of the most typical LC templates defined in the literature. Analogous to our hue templates, we use $W(P_i)$ to find the optimal $\epsilon^*_n$ for each template $LC_n, n=1 ... 6$, and the best fitting template $LC_n^*$.}

\rev{
Snapping colors to a template $LC_n$ requires finding the 2D line that fits best the LC distribution of the colors over a narrow hue band. To do that we minimize a weighted sum of all the perpendicular distances from each color to the axis of $LC_n$, weights are the same $W(P_i)$ from Subsection~\ref{sec:harmony_Harmonization: template fitting}. Specifically:
\begin{itemize}
\item For $LC_1$, the optimal position for the vertical axis after the optimization is $\epsilon_1^*=\sum W(P_i)*C(P_i)$. 
\item For $LC_2$, the optimal horizontal axis is $\epsilon_2^*=\sum W(P_i)*L(P_i)$.
\item For $LC_3$ and $LC_4$, we look for the axis pivoting from $0_{LC}= \{0,0\}$ and $1_{LC}= \{1,0\}$. We search for the axis rotation by brute force every $1^{\circ}$ to find the optimal $min(\epsilon_3^*, \epsilon_4^*)$
\item For $LC_5$, the diagonal line equation is $x-y-d=0$, where $x$ is $C$ and $y$ is $L$. Then optimal displacement $\epsilon_5^* = \sum W(P_i)*(C(P_i)-L(P_i))$
\item For $LC_6$, the line equation is $x+y-d=0$. Then optimal displacement $\epsilon_6^* = \sum W(P_i)*(C(P_i)+L(P_i))$.
\end{itemize}

For all templates, after line fitting we find the two extreme colors for the axis, and space the remaining ones evenly between those. 
}

\begin{figure}
	\centering
	\includegraphics[width=\columnwidth]{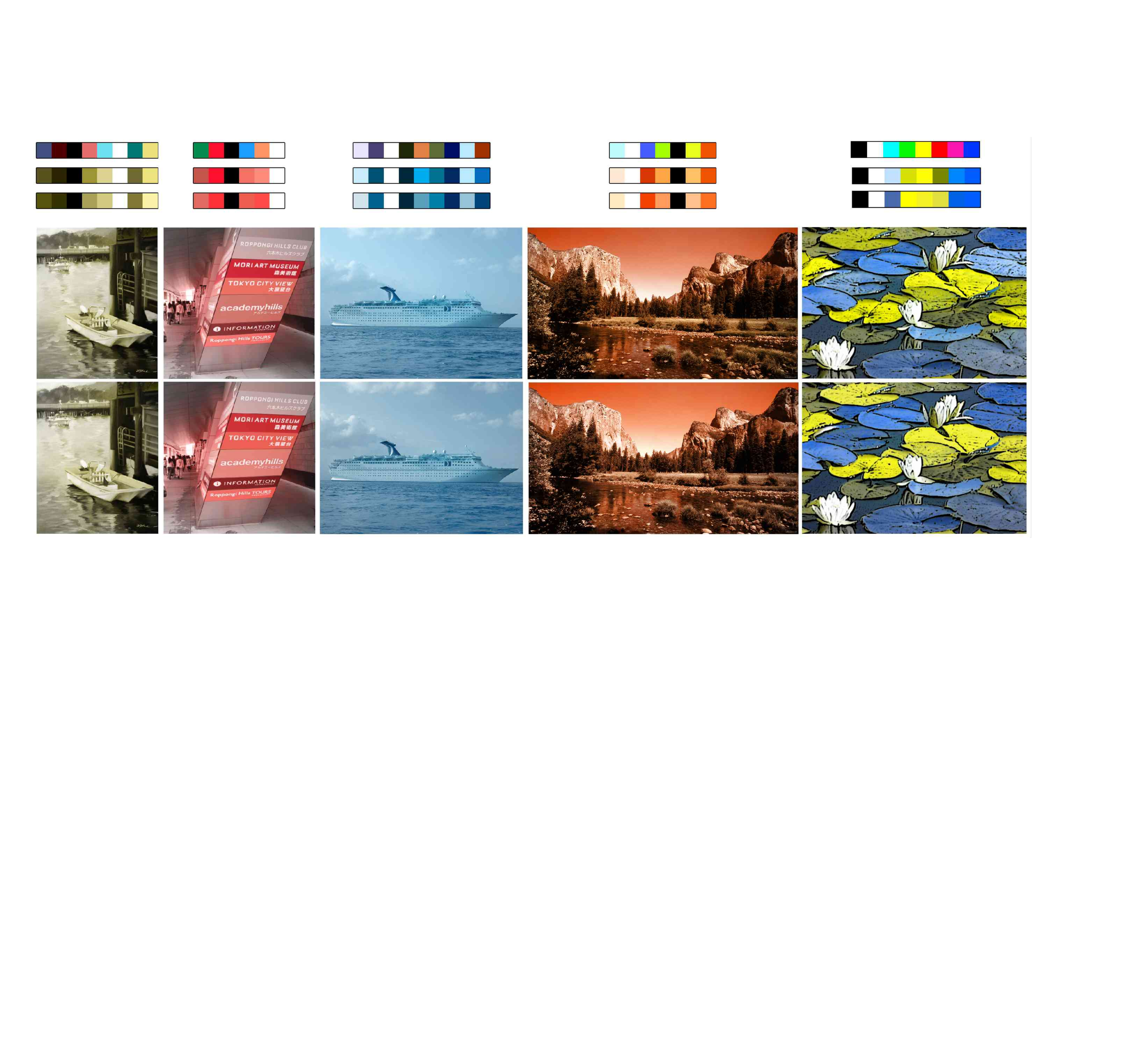}
	\caption{\rev{LC harmonization examples. From top to bottom: Extracted palette from input image. Palette hue harmonization for the \emph{monochromatic} and \emph{complementary} (last column only) template. Palette LC harmonizations as seen in Figure~\ref{fig:LC_templates}. Image hue harmonization. Image LC harmonization applied on top of the previous hue harmonization. Changes from LC harmonization are more apparent between the palettes, and more subtle on the recolored images. This is expected because $LC_n^*$ is applied to each image, producing more subtle results.}}
	\label{fig:LC_results}
\end{figure}

\rev{As can be seen, $LC_n$ are defined primarily for a single axis, and so they are directly applicable to monochromatic and analogous hue templates. For multi-axis templates, specific arrangements are described by Munsell~\cite{Birren1969} for complementary schemes, in terms of visual balance between the two axes, pivoting around a neutral point. We implement this idea by applying $LC_{3,4}$ pivoting around $N_{LC}=\{0.5,0\}$ for each axis. This approach can handle an arbitrary number of axis, although for palettes of optimal size, sometimes it is difficult to find more than one color per axis. Figure~\ref{fig:LC_results} shows examples of LC harmonization. It is worth mentioning that while our hue harmonization is always able to produce colorful results that preserve shading and contrast, harmonizing lightness and chroma may produced unwanted loss of contrast when enforcing templates other than the optimal $LC_n^*$.} 
	
% \paragraph{LC constraints}
% Constrains in luminance and chrome (LC) can create even more pleasing color combinations~\cite{tokumaru2002color,Moon44}.
% These can be enforced automatically like we do for our chroma and hue (Ch) templates, but they could also be explored interactively by the user.
%We implemented a simple GUI that helps the user to achieve LC arrangements like the ones described in classical color theory~\cite{tokumaru2002color,Moon44}. For images, we found changes in L or C tend to produce unappealing results because of the resulting effects over gradients and contrast.
% \jianchao{a figure to show a interactive editing GUI and some results?} \jianchao{currently we are interactively change LC. maybe Automatically adjust LC according to Hue change???} 
% \jose{depending on how it looks, we may not need this paragraph. I'm not sure it adds that much, so maybe move to discussion or future work.}

%After we harmonized palette colors' Hue channel by fitting some specific harmonization template, we can also edit Luminance and Chroma channel to enable image contrast adjustment according to modified Hue values. We also provide a convex hull based editing tool in LC space, where user can drag vertices to change palette's LC values interactively.
%\jianchao{a figure to show a interactive editing GUI and some results?} %\jianchao{currently we are interactively change LC. maybe Automatically adjust LC according to Hue change???}

\begin{figure}
	\centering
	\includegraphics[width=\columnwidth]{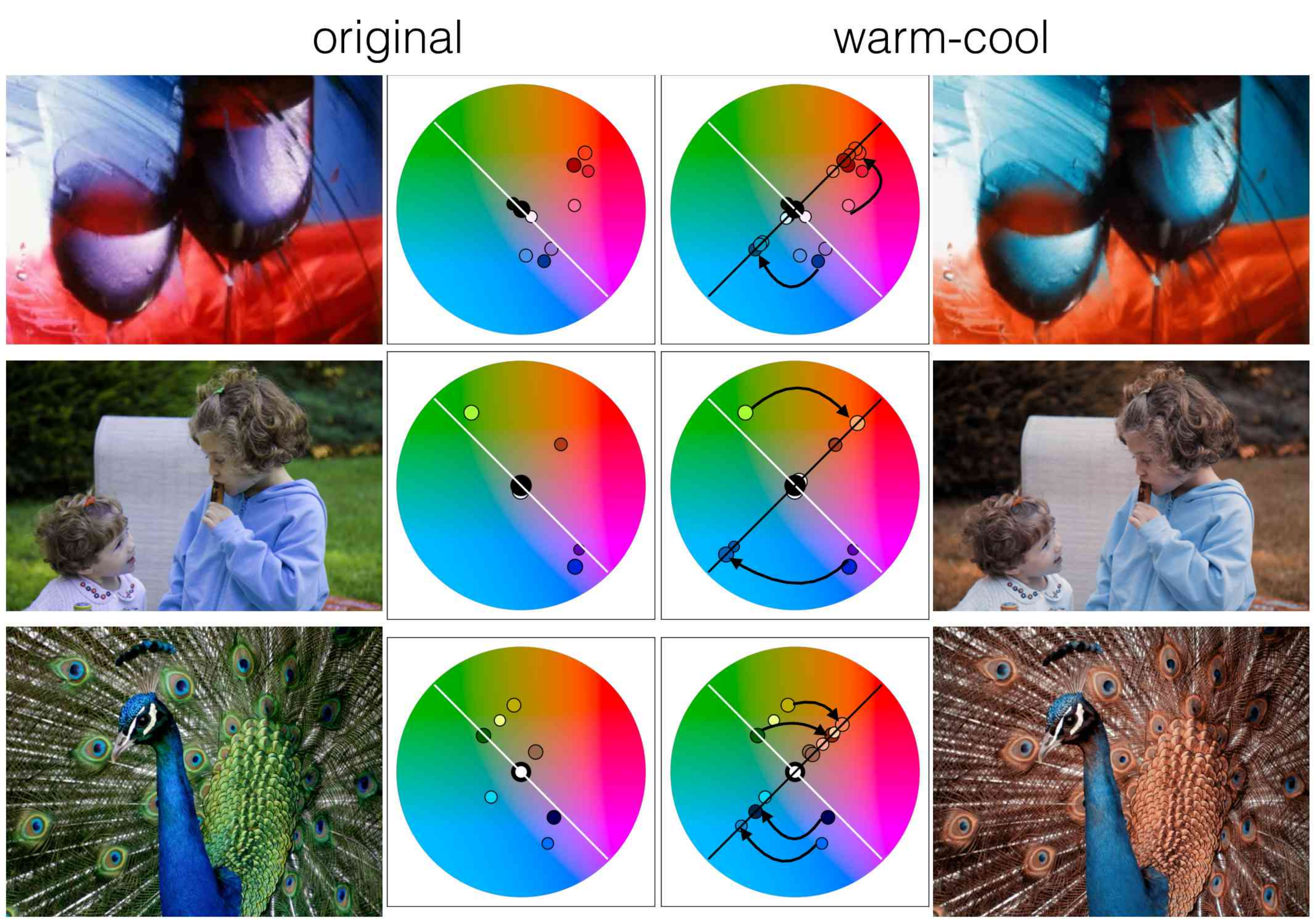}
	\caption{\revbegin Examples from the warm-cool contrast operator.}
	\label{fig:color_contrast}
\end{figure}

\paragraph{Color-based contrast}
\label{sec:harmony_Harmonization: constrast}
As part of his seminal work on color composition for design, \citet{itten1970elements} described additional pleasing color arrangements to create contrast.
In contrast with sector-based templates,
it is straightforward to model them with our axis-based representation. Here is the exhaustive list of Itten's additional contrasting color arrangements and how they fit into our framework:

\begin{itemize}
	\item Hue: \emph{Triad} template aligned with the RGB primaries. No need to solve for $\alpha^*$. 
	\item Light-dark: \emph{analogous} or \emph{monochrome} template, \rev{plus $LC_1$}.
	\item Complementary: same as \emph{complementary} \rev{hue} template.
	\item Simultaneous: \emph{complementary} template, plus the axis with the smaller overall $W$ scales down its chroma \rev{proportionally to $\beta$.}
	\item Saturation: \emph{analogous} or \emph{monochrome} template, \rev{plus $LC_2$.}
	\item Extension: solve for L so the total sum of \rev{$L(P_i)C(P_i)W(P_i)$} for each axis $j$ in $T_m^j(\alpha^*)$ is the same.
	\item Cold-warm: a \emph{complementary} template whose axis is aligned perpendicular to the cold-warm divide. \rev{The cold-warm divide is the complementary axis from red to cyan as seen in Figure~\ref{fig:color_contrast}}.
	
\end{itemize} 

% \jose{How many of them were actually implemented?}
% \jianchao{a figure to show results}

%% file: perceptual.tex
\revbegin
\section{Perceptual Study}
\label{sec:perceptual}

\begin{figure}
\centering
\includegraphics{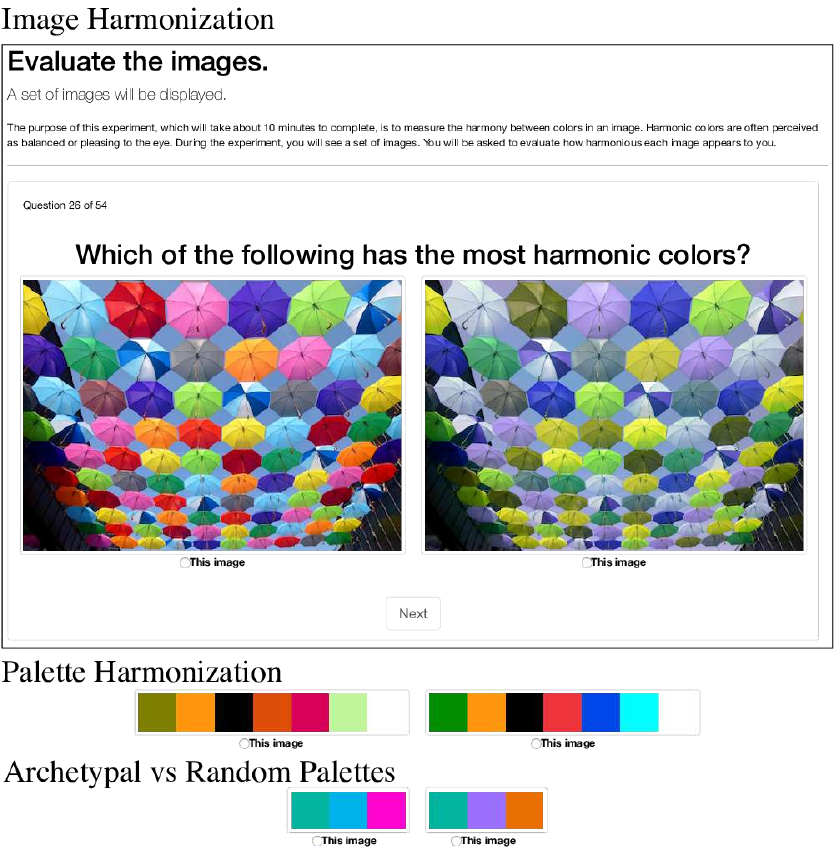}
\caption{\rev{Instructions and sample stimuli from our three perceptual experiments.
All stimuli can be seen in the
supplemental materials.}}
\label{fig:study_stimuli}
\end{figure}

We conducted a set of wide-ranging perceptual studies on harmonic colors and our harmonization algorithm.
$N=616$ participants took part in our studies with 31\% self-reporting as having some knowledge in color theory.
We obtained between $1000$ and $3000$ ratings per template, depending on the study.
In our first study, we performed an end-to-end evaluation of our image harmonization algorithm and \citet{Cohen-Or:2006:CH:1179352.1141933}.
To disentangle image content from color, we conducted a second evaluation
on our harmonized palettes alone.
Finally, to disentangle our algorithm from the percept of color harmony,
we conducted a study evaluating the perception of archetypal harmonic color schemes.

In our experiments, we avoided the use of Likert scales,
because the anchoring or range is unclear.
While a given harmonic scheme can be applied with varying strength
($\beta$ in Section~\ref{sec:harmony_Harmonization: template fitting}),
different harmonic schemes are incomparable.
If shown all harmonized images in a gallery, participants may develop anchors for the Likert scale between templates.
If shown harmonized images one-at-a-time in sequence, the same phenomenon would occur,
but the anchors would develop dynamically across the sequence.

Therefore, all of our experiments are based on 2-alternative forced-choice (2AFC) questions
\cite{cunningham2011experimental}.
Participants were shown two images and asked to choose which of two images has the most harmonic colors (Figure~\ref{fig:study_stimuli}).
The instructions explained that, ``Harmonic colors are often perceived as balanced or pleasing to the eye.''
In all experiments, a participant saw every stimulus (pair of images) twice. We used blockwise randomization so that, for each image, all stimuli were seen once before they were seen a second time.
We used rejection sampling to guarantee that no stimuli was seen twice back-to-back.
The initial left/right arrangement of the pair was random. For balance, the second time the pair was shown in the opposite arrangement.
We do not discard data from participants who answer inconsistently.
If a participant cannot decide, they are expected to choose randomly.

All stimuli and study data can be found in the supplemental materials.

\subsection{Image and Palette Harmonization}
\label{sec:perceptual:image_harmonization}

In our first experiment, we evaluated the output of image harmonization.
Each survey compared an unmodified image to various harmonization algorithms:
our monochromatic, complementary, single split, triad, double split, square, analogous,
and two LC harmonization algorithms (monochromatic and complementary),
and the output of \citet{Cohen-Or:2006:CH:1179352.1141933}.
For all algorithms, we compared the unmodified image to the harmonized.
For our harmonization output, we also compared the unmodified image
to the harmonization applied 50\% ($\beta = 0.5$), and the harmonization applied 50\% to the harmonization applied 100\% ($\beta= 1$).
We did not compare different templates directly.

% \paragraph{Experiment.}
We hypothesized that the harmonized images would be preferred, perhaps weakly,
by viewers.
We further hypothesized that this preference would vary by template,
and that the preference would decrease when applying
templates which lead to smaller changes in the output.
If the palette change to match the metric is small,
then the harmonized image may be indistinguishable from the original.
In 2AFC experiments, this causes participants to choose randomly,
so the preference tends towards 50/50.

We ran on our experiment on 25 images,
9 of which had output from \citet{Cohen-Or:2006:CH:1179352.1141933}.
We recruited $N=350$ participants via Amazon Mechanical Turk,
29\% of whom reported having some knowledge in color theory. Individuals with impaired color vision were asked
not to participate in the study.
We sought 1000 ratings per template in order
to detect an effect size of approximately 55\% with a factor-of-10 correction
for multiple comparisons (\v{S}id\'{a}k or Bonferroni) due to our 10 harmonization algorithms.
To obtain 1000 or more ratings per pair of images, we obtained ratings from 20 individuals for each of the harmonizations of the 16 images without
\citet{Cohen-Or:2006:CH:1179352.1141933}'s output,
and from 60 individuals for each of the harmonizations of each of the 9 images with
\citet{Cohen-Or:2006:CH:1179352.1141933}'s output.
(Each individual rated each pair twice.)

\begin{figure}
\centering
\includegraphics[width=\linewidth]{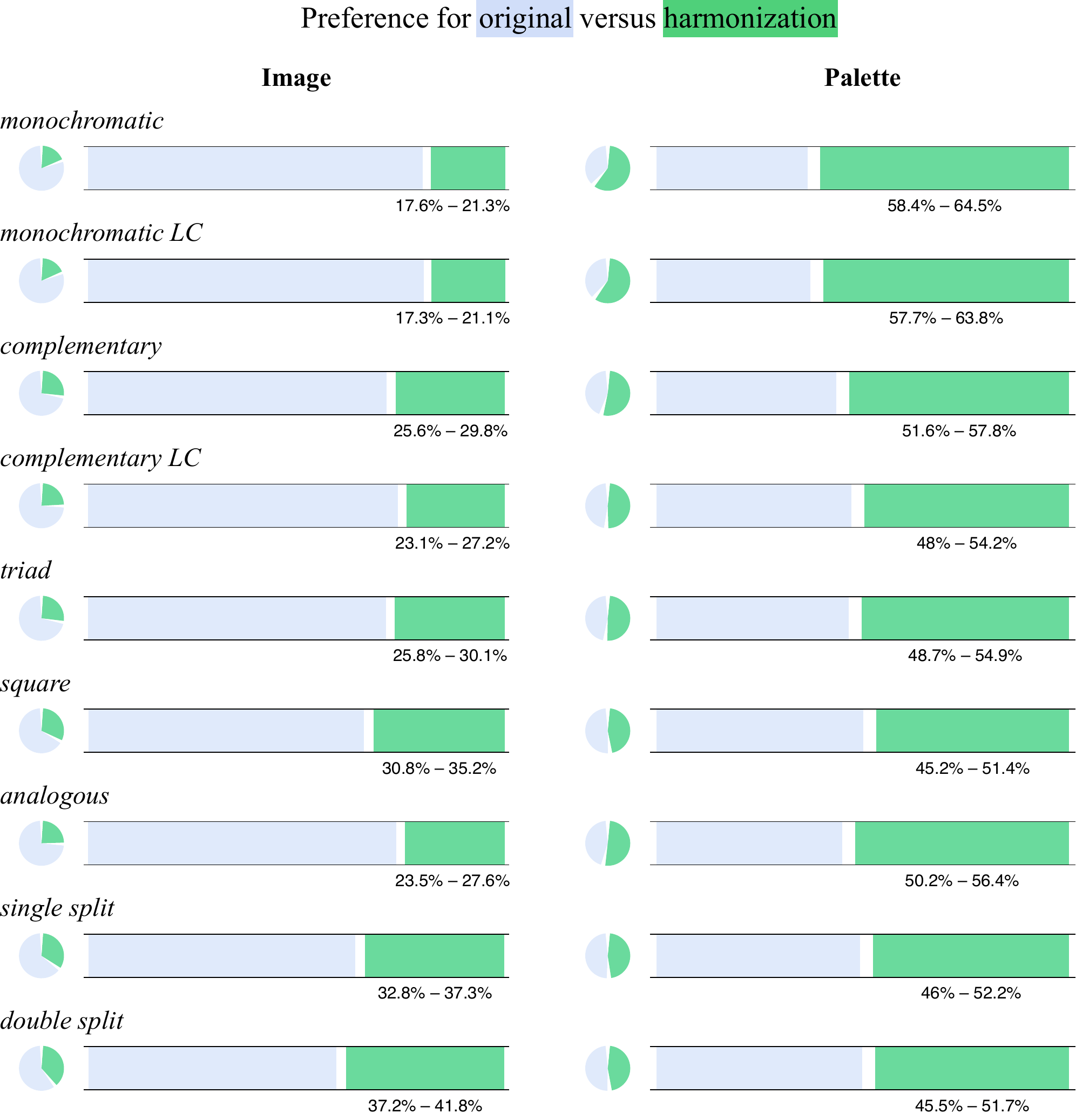}
\caption{Across the set of all 25 images, the harmonized images for any given template were not judged to be more harmonic than original image,
whereas the palettes often were.
The colorful intervals have $95\%$ confidence.
Participants with knowledge in color theory
preferred our image and palette harmonizations, on average,
by an additional 3.7\% and 4.5\%, respectively.
}
\label{fig:perceptual_image_harmonization_all_templates}
\end{figure}

\begin{figure}
\centering
\includegraphics[width=.7\linewidth]{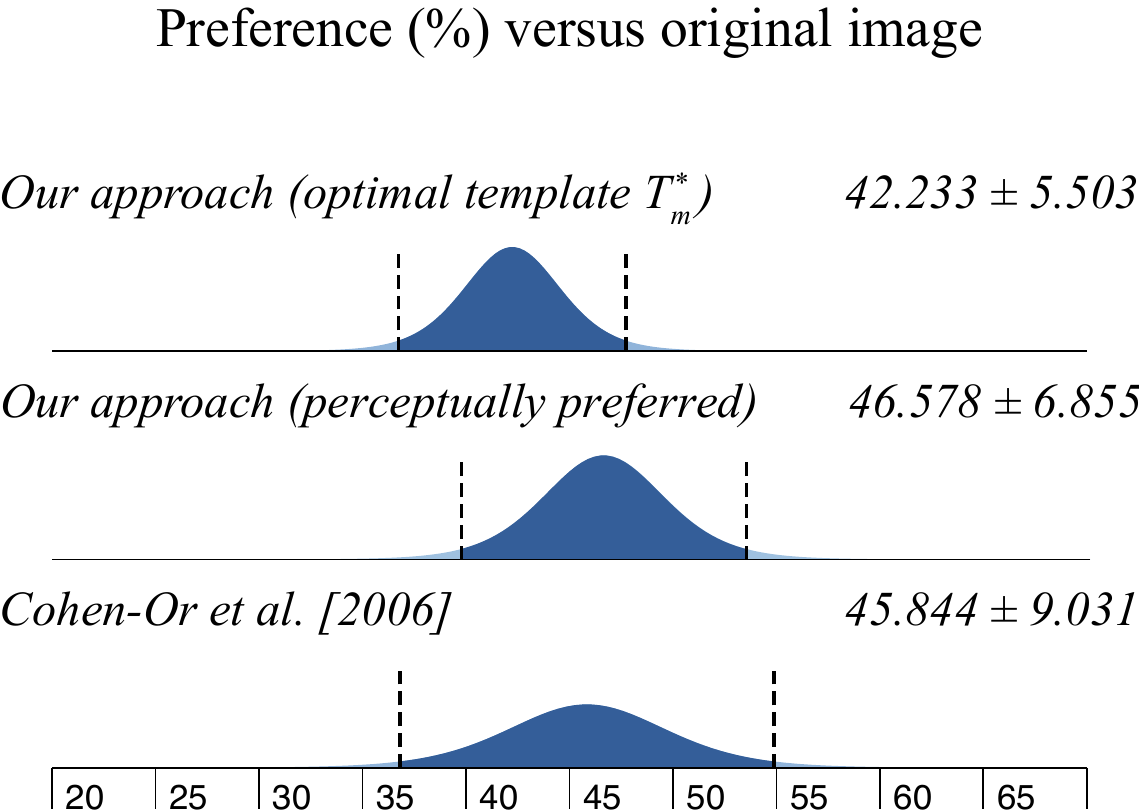}
\caption{Across a set of 9 images used in \protect\citet{Cohen-Or:2006:CH:1179352.1141933},
we compare the rate of preference versus unmodified images
for \protect\citet{Cohen-Or:2006:CH:1179352.1141933}'s technique
and for the ``best'' harmonized output of our technique in two senses:
using the template chosen by our optimal metric $T_m^*$ (Equation~\ref{eq:harmony:optimal_template})
and using the template most liked by participants in the perceptual study.
The differences between these distributions are not statistically significant.
}
\label{fig:perceptual_vs_cohenor}
\end{figure}

The most notable observation about this first study is that participants
overall preferred the original images to harmonizations
and a preference for $\beta=0.5$ to $\beta=1$ (Figure~\ref{fig:perceptual_image_harmonization_all_templates}, left).
% We performed a second study to determine whether 
%
While any given harmonization
was not preferred to the original across all images,
there was substantial variation per-image.
For example, an analogous template fared better on some images versus others.
Participants with knowledge about color theory had a statistically significant ($p \ll 0.001$) stronger preference for harmonized images (3.7\% overall).

In addition to our 9 harmonization templates,
we also evaluated \citet{Cohen-Or:2006:CH:1179352.1141933}'s harmonization result
on a subset of 9 images.
Because we only have \citet{Cohen-Or:2006:CH:1179352.1141933}'s optimal harmonization
result, we compared preference rates to our automatically-chosen
optimal harmonization $T_m^*$ (Equation~\ref{eq:harmony:optimal_template})
and to the harmonization template most preferred by participants in the perceptual study (Figure~\ref{fig:perceptual_vs_cohenor}).

Harmonizing the colors of natural images was noted as a limitation by
\citet{Cohen-Or:2006:CH:1179352.1141933} due to our expectations.
In their output, they used masks to avoid, for example, affecting human skin.
(We do not.)
However, several of the images in our study were not natural images
with no apparent effect on ratings (for our technique and Cohen-Or's).
To investigate whether the image content was biasing partipants' perception,
we performed a second perceptual study that repeated the experiment,
replacing every image with its palette.
For this study, we recruited $N=200$ participants via Amazon Mechanical Turk,
32.5\% of whom reported having some knowledge in color theory.
Because \citet{Cohen-Or:2006:CH:1179352.1141933} is not palette-based,
we omitted it from the study since there are no before/after palettes to evaluate.
Therefore, to obtain 1000 ratings per comparison, we obtained ratings from 20 individuals for each of the harmonizations for all 25 images.

In this experiment, the harmonizations were judged significantly better than when
displaying images (Figure~\ref{fig:perceptual_image_harmonization_all_templates}, left).
The harmonizations on average were preferred to the original palettes ($\chi^2 = 37, p \ll 0.001$).
Our monochromatic ($\chi^2 = 53, p \ll 0.001$), monochromatic LC ($\chi^2 = 47, p \ll 0.001$), and complementary ($\chi^2 = 9, p=0.027$ with factor-of-nine Bonferroni correction) harmonizations
produced palettes preferred to the originals.
Participants with knowledge about color theory had a statistically significant ($\chi^2 = 4, p \ll 0.001$) stronger preference for harmonized palettes (4.5\% on average).
Among knowledgeable participants,
each template's harmonized palettes were preferred to the originals $>50\%$ of the time.
The same three templates (monochromatic, monochromatic LC, and complementary) were preferred with statistical significance;
the power of our study when restricted to knowledgable participants (344 ratings per template) had insufficient power to conclude whether the preference for additional specific templates was significantly different than chance.

We expected all of our harmonized results to be judged more harmonic than the input.
Since many harmonizations of the same image or palette were shown to participants,
there may have been a familiarity bias towards the more common original image or palette.
To eliminate this as well as any biases stemming from the complexity of the image palettes,
we performed an additional study.

% \todo{Something about Mechanical Turk population \cite{difallah2018demographics}.}

% Future work: Evaluate more steps to detect JND and fit a psychometric function to color harmony appreciation \cite{cunningham2011experimental}.

\subsection{Perception of Archetypal Color Harmony}
\label{sec:perceptual:archetype}

\begin{figure}
\centering
\includegraphics[width=\linewidth]{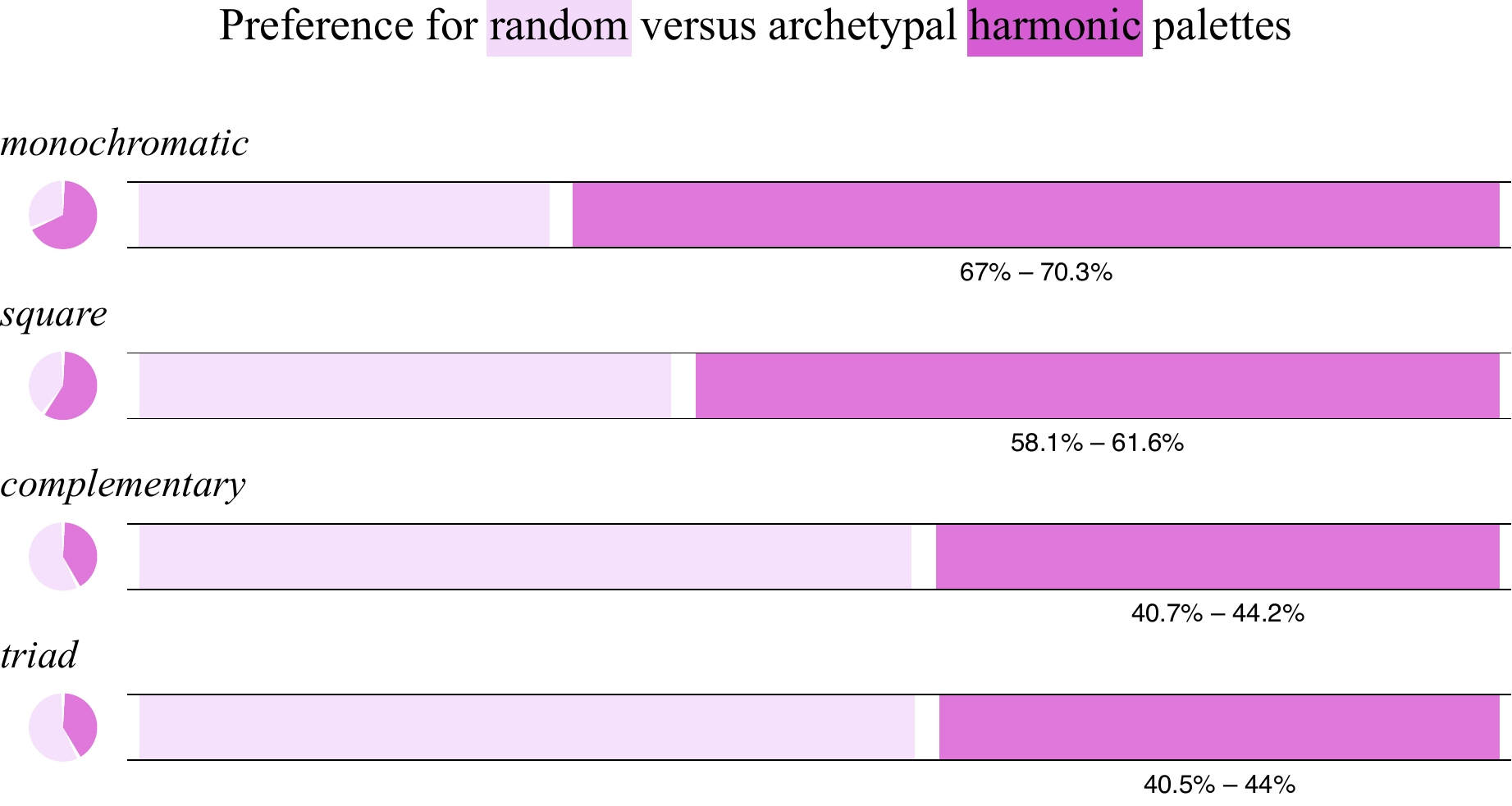}
\caption{Monochromatic and square archetypal harmonic color templates were perceived to be more harmonic than random palettes,
whereas complementary and triad palettes were not.
Participants with and without knowledge of color theory exhibited the same preferences.
}
\label{fig:perceptual_archetype}
\end{figure}

To evaluate whether color harmony can be perceived in an archetypal setting,
we evaluated the following basic templates in a controlled manner:
monochromatic, complementary, triad, and square.
We generated random monochromatic, complementary, triad, and square palettes with one, one, two, and three colors, respectively.
For the complementary, triad, and square palettes,
we randomly generated one
color and then spaced the rest $180^\circ$, $120^\circ$, and $90^\circ$ around the
color wheel. We used Lch color-space, which is a cylindrical parameterization of Lab color-space. All colors had luminance 60 and chroma 100.
The monochromatic template consisted of two colors. The first color had luminance 60, chroma 100, and randomly chosen hue. The second color was obtained by scaling
the luminance and chroma of the first by two random factors in the range $[0.5,0.75]$.
We generated 15 palettes for each of the four categories.

To eliminate any familiarity bias, each of the 60 palettes were paired with a unique, random palette. Each palette was shown exactly twice with the same pairing.
The random palettes shared the first color with their paired harmonic palette.
For random palettes, we obtained the remaining color(s) by randomly sampling hues.
We ordered the remaining colors according to their hue relative to the first color around the color wheel.
For random palettes paired with complementary, triad, or square palettes, luminance and chroma were uniform across the entire palette.
For random palettes paired with monochromatic palettes, the remaining color shared the same luminance and chroma as the second color in the monochromatic palette.
We used rejection sampling to ensure that we didn't accidentally generate a palette fitting one of the harmonic templates.
(No two colors can be less than 23 units apart in Lab space, which is 10 times the just noticeable difference.)

We recruited $N=100$ participants via Amazon Mechanical Turk,
38\% of whom reported having some knowledge in color theory.
Each participant saw all pairs of palettes with the aforementioned randomization scheme (120 questions) and presentation (Figure~\ref{fig:study_stimuli}).
Each of the four templates therefore received $3000$ evaluations versus a random palette.

The monochromatic and square templates were perceived to be significantly more harmonic
than random palettes (Figure~\ref{fig:perceptual_archetype}).
However, random templates were perceived as more harmonic than complementary and triad templates.
In this study, participants with knowledge about color theory did not significantly differ in their judgments from participants without knowledge.
We conjecture that the complementary and triad templates created the most contrast, which
may have been the primary phenomena participants considered when
evaluating harmony; in other words, strong contrast was perceived to be disharmonious.
This experiment suggests that perceptual uniformity in hue intervals may not be consciously perceived as ``balanced or pleasing to the eye.''

\revend

%% file: video.tex
\revbegin
\section{Video Harmonization}
\label{sec:harmony_video}
Our methods can naturally extend to video by simply applying our image decomposition and harmonization on each frame independently. In this case, given the properties of our extracted palettes, we first compute a global palette for each sequence of frames, aiming at a more coherent layer decomposition without additional processing beyond the proposed framework.  
%Our RGBXY hull methods will automatically guarantee temporal coherence of the final harmonized video, without need to solve optimization for temporal coherence, as did by all previous video editing papers. 
We describe the overall pipeline in Algorithm~\ref{algo:video_pipeline}.

\begin{algorithm}
\revbegin
\DontPrintSemicolon % Some LaTeX compilers require you to use \dontprintsemicolon instead
\KwIn{Original video frames $F$, frame number $N$ and harmonization template $T$}
\KwOut{Harmonized video frames $H$.}
// Aggregate dense convex hull vertices of each frame $F_i$ \;
$i \gets 0$\;
$I \gets $ \o \;
\While{$i<N$}{
	$I \gets I \cup$ ConvexHull ($\mathbf{F_i}$) \;
	$i \gets i+1$\;
	}
// Extract the global palette of the video sequence\;
P$_{\textit{original}}\gets$ Simplify( ConvexHull( $\mathbf{I}$ ) )\;

// Get averaged mixing weights for the global palette \;
$i \gets 0$\;
$W_{\textit{sum}} \gets \mathbf{0}$ \;
\While{$i<N$}{
    $W_i \gets $ LayerDecomposition( $F_i$ , P$_{\textit{original}}$) \;
    $W_{\textit{sum}} \gets W_{\textit{sum}}+ W_i$  \;
	$i \gets i+1$\;
	}

$W_{\textit{avg}} \gets W_{\textit{sum}} / N$ \;
// Harmonize the global palette\;
P$_{\textit{harmonized}} \gets$ Harmonize(P$_{\textit{original}}$, $W_{\textit{avg}}$, $T$)\;
// Recolor frames with the harmonized palette\;
$i \gets 0$\;
\While{$i<N$}{
	$H_i \gets W_i \cdot$ P$_{\textit{harmonized}}$ \;
	$i \gets i+1$\;
	}
\Return{$H$}\;
\caption{Our proposed video harmonization process.}
\label{algo:video_pipeline}
\end{algorithm}

\rev{We show examples of video harmonization in Figure~\ref{fig:video_harmonization_example}. Videos can be found in the supplementary material.}

\begin{figure}
	\centering
	\includegraphics[width=\columnwidth]{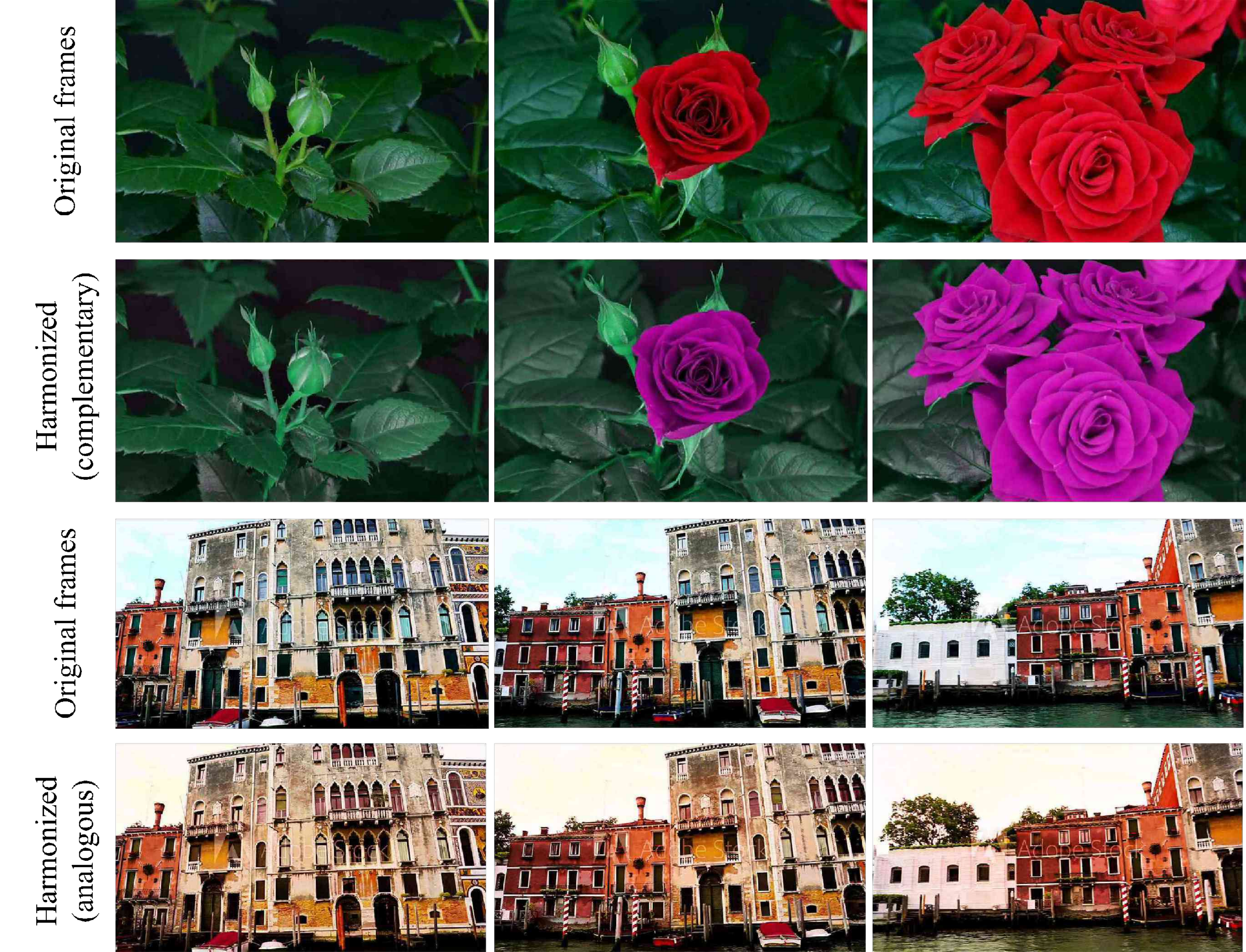}
	\caption{\rev{Video harmonization results. First and last frames are shown for each sequence, along an intermediate one to appreciate changes in color palette over time. Consistent harmonization is achieved across all of them.}}
	\label{fig:video_harmonization_example}
\end{figure}

\revend

%% file: transfer.tex
\section{Color Transfer}
\label{sec:harmony_ColorTransfer}

Our palette extraction, image decomposition, and harmonic templates enable
new approaches to color transfer.
Harmonic templates carry important information about the color distribution in a palette or an image. We propose to transfer that information between palettes and images.
% From Section~\ref{sec:harmony_Harmonization}, we can see how h

%We use optimal template fitting results as intermediate matching tool to transfer reference image palette colors to input image palette colors. We provide two options for transferring colors, as shown in Figure XX.

\begin{figure}
	\centering
	\includegraphics[width=\columnwidth]{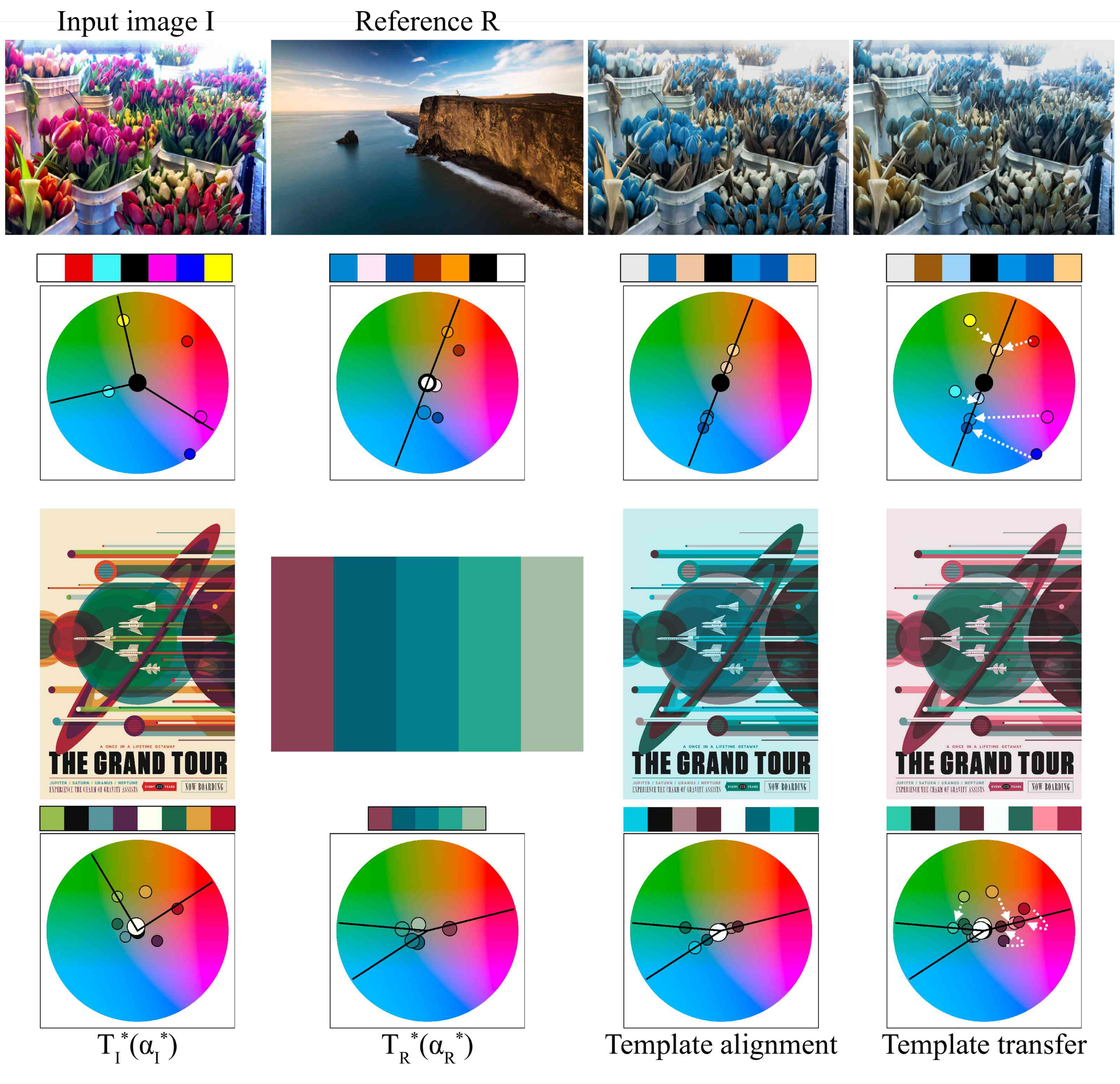}
	\caption{\rev{Comparison between our two template-based color transfer methods. Third column shows how aligning the input and reference templates, $T_I^*(\alpha_I^*)$ and $T_R^*(\alpha_R^*)$ respectively, transfers better the overall color proportions, something that tends to work better for content without critical color semantics (bottom example). On the other hand, template transfer (rightmost column), preserves better the original colors.}}
	\label{fig:color_transfer_template}
\end{figure}

\paragraph{Template alignment} Given an input image $I$ and a reference image $R$, we already know how to extract their palettes $P^I$ and $P^R$, and estimate their optimal templates, $T_I(\alpha_I^*)$ and $T_R(\alpha_R^*)$.
After the fitting, we can compute the weight of each axis of the template as the sum of the weights of each color $W(P_i)$ assigned to it. With this, we have an estimate of the main axis for each template---the one with the greatest influence on the image. This simple procedure helps to establish a straightfoward match between palettes, something we can leverage to find the global rotation $\gamma$ that aligns $T_I$ with $T_R$. Next, we apply $\gamma$ to globally rotate the colors of $P^I$ and then we harmonize them with the target's template $T_R(\alpha_R^*)$ with $\beta = 1$.
 % to it \jose{this calls for the missing equation in Sec 4. Probably a function $H(P, T(\alpha), \omega)$ that enforces $T(\alpha)$ over $P$, with a strength $\omega$ }.
 This method achieves results where $I$ is recolored so it is harmonic with respect to $R$, taking into account the overall relevance of each color of the palette. Figure~\ref{fig:color_transfer_template} shows results of this approach.
 We found that this method is good for matching dominant colors,
which works better for content without real reference colors
(e.g.\ graphics design or man-made objects).

%We first find optimal templates $T_r$ and $T_i$ for reference image palette $P_r$ and input image palette $P_i$, as described in section~\ref{sec:harmony_Harmonization: template fitting}. After template fitting, we know the matching correspondence between palette colors and optimal template's axises, we accumulate the pixel weights (mentioned in equation~\ref{eq:harmony:palette_tempalte_distance}) of these colors onto their corresponding axises. Then for each axis, we will have a accumulated weight. We will denote the axis with largest weight value as the main axis of that template. And we will align main axis of input image optimal template $T_i$ onto main axis of the reference image's optimal template $T_r$, and all input palette colors will globally rotate with template together to some where in HC wheel, denote it as $P_i^{(1)}$. After rotation, we will perform harmonization template fitting step again for current rotated input palette colors $P_i^{(1)}$ and reference image's optimal template $T_r$, then we will obtain final harmonized input palette colors $P_i^*$, which is a similar color pattern as reference image's palette colors pattern.

\paragraph{Template transfer}  When the final results should  preserve better the original colors, a more conservative method can be formulated. In this case, we
harmonize the input image colors $P^I$ directly to the
the best-fitting template for the reference image $T_R(\alpha_R^*)$, without any global rotation.
We match palette colors to template axes according to equation~\ref{eq:harmony:palette_tempalte_distance}.
%so it can be fitted directly over $P^I$, without taking into account the influence each color on the image.
%
% \bigbreak
%
After changing the hues of $P^I$ with any of the proposed methods, we attempt to match lightness and chroma between palettes by scaling the lightness and chroma of
each palette color to that the average $L/C$ of the input and reference palette colors match.

\begin{figure*}
	\centering
	\includegraphics[width=\textwidth]{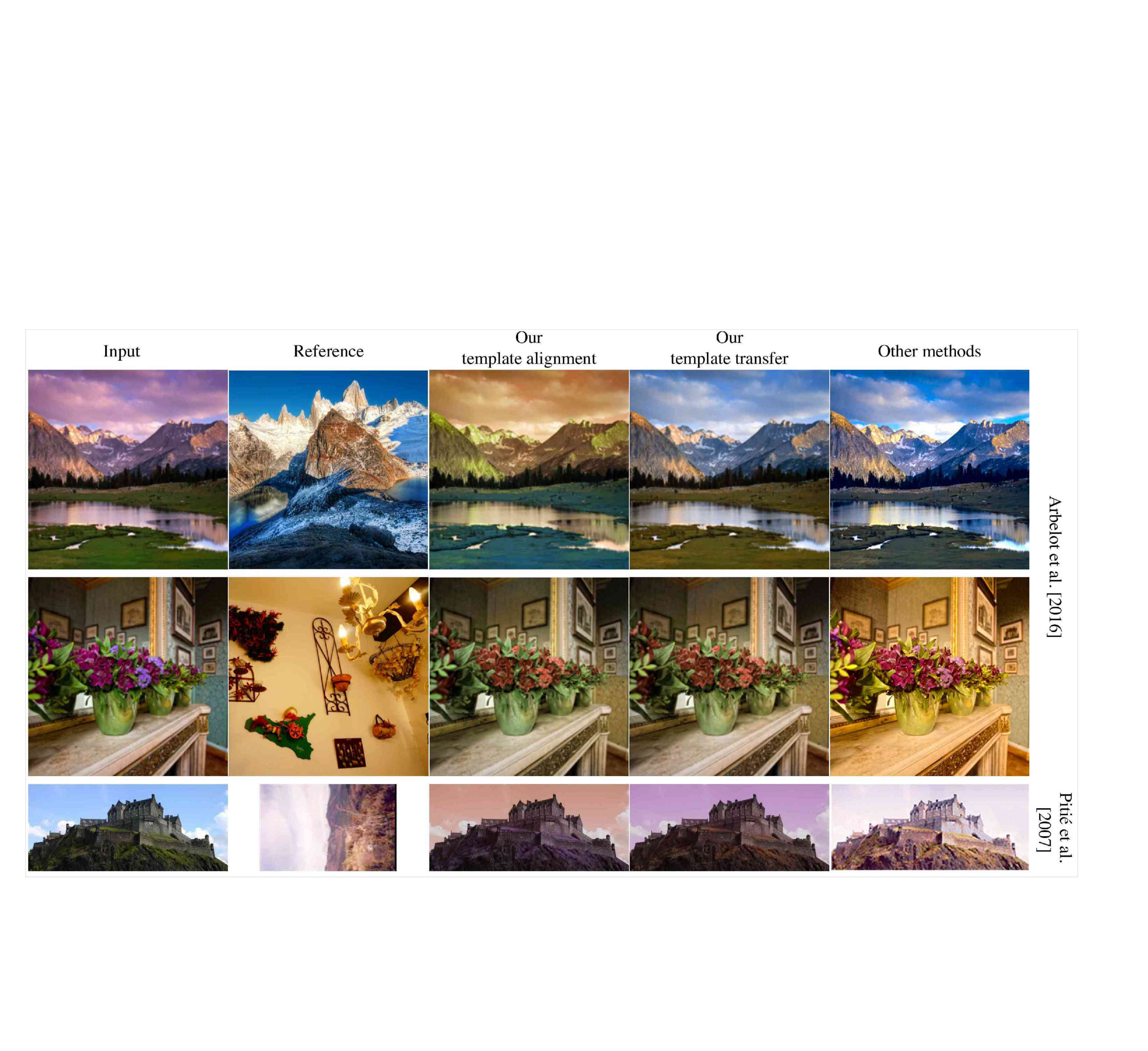}
	\caption{Color transfer comparison with some previous works:~\citet{Arbelot2016} (first and second rows) and \citet{Pitie2007} (bottom row). From left to right: input image, reference, template alignment, template transfer and the related works. Our methods provide some results closer to~\cite{Arbelot2016}, especially template transfer (column 4). Compared to~\cite{Pitie2007}, our transfers do not capture the overall tone that well, but produce usable stylized results.}
	\label{fig:color_transfer_previous}
\end{figure*}

Figure~\ref{fig:color_transfer_template} shows results from this method.
We compare both color transfer approaches with previous work in Figure~\ref{fig:color_transfer_previous}.

%% file: conclusion.tex
\section{Conclusion}
\label{sec:harmony_conclusion}

We have presented a very efficient, intuitive and capable framework for color composition. It allows us to formulate previous and novel approaches to color harmonization and color transfer with very robust results. Our palette manipulations can be plugged into any palette-based system. Our image decomposition can be used generally by artists for manual editing or in other algorithms.
\rev{Our large-scale perceptual study provides important data and insights into the perception
of color harmony.}
% While we show plenty of convincing results, there are several aspects that we would like to keep exploring.

\subsection{Limitations}
\label{sec:harmony_conclusion:limitation}

During our performance tests for the image decomposition, we found isolated cases where the computation of the 5D convex hull takes somewhat longer than usual. We believe it is due to very specific color distributions ($3$ out of $170$ tested images), but we would like to study the phenomenon in more depth.

There are also cases for the templated color transfer where the input palette tries to match a reference palette with a higher number of axes. This is usually a case of colorization (adding more colors than the existing ones) that we currently handle with varying success depending on the input color palette. These cases may need more elaborate formulations for the transfer. %Our non-templated transfers help alleviate the problem.

Because there is not a universal color theory, the concepts we leverage for our methods may not work for everybody. In fact, we already saw clear differences in our results with respect to previous work, even building on top of comparable foundations.
\rev{Our perceptual study has revealed potential problems
with the percept of color harmony that affect all work on color harmony.}
This exposes the need for \rev{additional} perceptual studies evaluating the perceived quality of results from different algorithms.
This also exposes the need for intuitive frameworks like ours, enabling users
to use and interact with color harmony, despite only passing familiarity,
\rev{so that they might find something indeed balanced and pleasing to the eye}.

%Our algorithms have some limitations in some steps. One is that convex hull on 5D data is sometimes time consuming. It is hard to predict the time, since it is not that positively related with image size. This is the bottleneck in current fast RGBXY method. 
%Second is that when we transfer color between two images, current template based palette color transfer is not good for the case that a simpler template mapping to a complexer template, which is a colorization problem. Directly matching two palette will sometimes work well for this situation, but sometimes not.

%There are three isolated red dots for RGBXY method in figure, the dominant time of these two examples falls into 5D convex hull computation step. Their RGBXY data distribution touches bad case of convex hull computation. However, only 3 of them are bad for all 170 images we tried.

\subsection{Future work}
\label{sec:harmony_conclusion:future}
\paragraph{Image decomposition}
Inspired by~\citet{Lin2017}, we wish to explore the use of superpixels to see if we are able to achieve greater speed ups.
%We also plan to explore replacing the current Delaunay barycentric coordinates with Mean Value Coordinates, as they do not require a Delaunay tessellation and may have additional smoothness benefits.
We also wish to explore parallel and approximate convex hull algorithms.
%  We may sample RGBXY 5D data to make 5D convex hull computation faster~\jose{???}. 

%\paragraph{Harmonization} Apart from our current chrome and hue (Ch) harmonic templates, we want to explore full 3D LCh templates for more complex palettes and compositions~\cite{tokumaru2002color,Moon44}. While altering lightness and chroma can have undesired effects on images (gradients, contrast), we believe this can lead to more varied color palettes and themes.

\paragraph{Other color-aware applications} We believe that our templates may carry semantic structure that we would like to keep exploring in the future. Among others, we believe this can enable higher level and more intuitive image search algorithms, where images or palettes can be used transparently to retrieve other images and color themes for design.
%We also also plan to extend our framework to video, exploring the spatial-temporal coherence of our templates, to potentially provide more robust color grading methods.
% We would also like to explore temporally and spatially varying color palettes.

%We may also want to explore 3D harmonization template for image harmonization. 
%We can extend our current image harmonization and color transferring framework to videos, by exploring the spatial-temporal palette relationship of each frame and the temporal coherence of recolored video frames. 

%Specially for enhancement of color palettes, adding constrains in LC can create even more pleasing color combinations. These can be enforced automatically like we do for the Ch case, but can also be explored manually by the user. We implemented a simple GUI that helps the user to achieve LC arrangements like the ones described in classical color theory~\cite{tokumaru2002color,Moon44}. For images, we found changes in L or C tend to produce unappealing results because of the resulting effects over gradients and contrast.
%\jianchao{a figure to show a interactive editing GUI and some results?} \jianchao{currently we are interactively change LC. maybe Automatically adjust LC according to Hue change???} 
%\jose{depending on how it looks, we may not need this paragraph. I'm not sure it adds that much, so maybe move to discussion or future work.}